\documentclass[draft]{agujournal2019}
\usepackage{url} 
\usepackage{lineno}
\usepackage[finalnew]{trackchanges} 
\usepackage{soul}
\usepackage{bm}
\draftfalse

\journalname{Radio Science}

\begin{document}

%
%


\title{Calibration of an SKA-Low prototype station using holographic techniques}

%
%

\authors{
Jishnu N. Thekkeppattu\affil{1},
Randall B. Wayth\affil{1},
Marcin Soko\l{}owski\affil{1}}

\affiliation{1}{International Centre for Radio Astronomy Research (ICRAR), Curtin University, Bentley, WA 6102, Australia}

\correspondingauthor{Jishnu N. Thekkeppattu}{j.thekkeppattu@curtin.edu.au (JNT)}




\begin{keypoints}
\item A framework for phased array holography using tensors is presented
\item Multiple holographic techniques are unified.
\item Self-holography and cross-holography of AAVS2 are performed and the results compared.
\end{keypoints}

%
%

%
%


\begin{abstract}
Performance of digitally beamformed phased arrays relies on accurate calibration of the array by obtaining gains of each antenna in the array. The stations of the Square Kilometre Array - Low (SKA-Low) are such digital arrays, where the station calibration is currently performed using conventional interferometric techniques. An alternative calibration technique similar to holography of dish based telescopes has been suggested in the past. In this paper, we develop a novel mathematical framework for holography employing tensors, which are multi-way data structures. Self-holography using a reference beam formed with the station under test itself and cross-holography using a different station to obtain the reference beam are unified under the same formalism. Besides, the relation between the two apparently distinct holographic approaches in the literature for phased arrays is shown, and we show that under certain conditions the two methods yield the same results. We test the various holographic techniques on an SKA-Low prototype station AAVS2 with the Sun as the calibrator. We perform self-holography of AAVS2 and cross-holography with simultaneous observations carried out with another station EDA2. We find the results from the holographic techniques to be consistent among themselves as well as with a more conventional calibration technique.
\end{abstract}


\section{Introduction}
The Square Kilometre Array (SKA) is an upcoming radio telescope, with its low frequency section (SKA-Low) being built in the Inyarrimanha Ilgari Bundara (Murchison Radio-astronomy Observatory) in Western Australia. The telescope will consist of 512 phased arrays called stations, with each station having 256 closed spaced dual-polarisation log-periodic antennas \cite{9107113}. The individual antenna voltages are digitised and added in the back-end firmware with appropriate beamweights to steer, and potentially shape, the beam in real-time. For accurate beamforming, the complex receive path gain of each antenna in a station must be accurately known. Obtaining the gain of each antenna in a station and its associated signal chain, called station calibration, is currently carried out on SKA-Low prototype stations Aperture Array Verification System 2 (AAVS2, \citeA{2020SPIE11445E..89V}) and Engineering Development Array 2 (EDA2, \citeA{2022JATIS...8a1010W} with mid-day observations of the Sun collecting intra-station visibilities. This is followed by conventional gain calibration techniques such as \texttt{mfcal} task implemented in the \texttt{miriad} radio astronomy package \cite{1995ASPC...77..433S} to derive the gains \cite{2021A&A...655A...5B,2021PASA...38...23S, 2022JATIS...8a1010W, 2022JATIS...8a1014M}.

However, station calibration using visibilities makes use of only the intra-station baselines from a station of $\sim$ 35m diameter. Even with a multi-station telescope, the interferometric station calibration would still be unable to take advantage of the other stations, as visibilities between individual antennas belonging to different stations are difficult to compute and store. As an alternative to \add{conventional} interferometric calibration of phased arrays, holographic techniques have been suggested. \remove{An attractive feature of holography is that only $\mathcal{O}(N)$ correlations between the antennas and a reference beam have to be calculated instead of $\mathcal{O}(N^2)$ antenna-antenna correlations, thereby vastly reducing the number of computations required.} Variants of holographic techniques have been successfully applied to calibrate a variety of phased array telescopes such as LOFAR \cite{2020A&A...635A.207S} and EDA2 \cite{2021RaSc...5607171K}. Holography has also been proposed as a potential calibration method for the mid-frequency aperture arrays (MFAA) for the second phase of the SKA \cite{2022JATIS...8a1008W}. \add{An attractive feature of holography is that it does not require the array covariance matrix, which can be beneficial for large arrays such as the MFAA \protect{\cite{2022JATIS...8a1008W}}}\add{, although in this paper we show that holography can be carried out with such a covariance matrix, if available.} Conventional holography of dish antennas uses two distinct antennas with one acting as a reference \cite{1977MNRAS.178..539S}. However, in the context of all-digital phased arrays, holographic techniques have been developed predominantly for self-holography in which the reference beam is provided by the same phased array. 

There are two distinct self-holographic techniques for phased arrays in the literature. The first technique, described in \citeA{8065418}, correlates each individual antenna with the reference beam. As shown in the same work, these "measured correlations" are then proportional to the receive path gains and can be solved for, if the signal to noise ratio (SNR) is high. \citeA{9369030} discusses the calibratability of phased arrays via this approach, where the effects of interfering sources are also discussed. They also apply self-holography to multiple telescopes. We will henceforth refer to this technique as the "Antenna-Correlation" approach in this paper, as the individual antennas are correlated with the reference beam. The second technique for self-holography, described in \citeA{2021RaSc...5607171K}, forms a reference beam and correlates it with voltage beams over a raster scan pattern on the sky. This method is closer to the more conventional holographic technique typically applied to dish antennas. The correlation of the beams over the sky results in "Beam-Correlation", a matrix that is subsequently Fourier transformed to obtain the complex aperture image. The output of this technique is an image with the antenna positions weighted by their relative complex gains, from which the individual receive path gains can be extracted. We will henceforth refer to this technique as the "Beam-Correlation" approach in this paper. Regardless of the chosen approach, implementation of \add{self-}holography for phased arrays \add{hitherto} required collection of individual antenna voltages. 

For both Antenna-Correlation and Beam-Correlation techniques, replacing the self-generated reference beam with a beam obtained from a physically separate antenna system leads to cross-holography. Thus, four different phased array holographic techniques are possible, which are listed in Table.\ref{Tab:holo_compare}.

\begin{table}[h]
\label{Tab:holo_compare}
\caption{List of holographic techniques}
\centering
\begin{tabular}{l l l}
\hline
Holographic technique                   & Reference beam source    & Output   \\ \hline
Self-holography with Beam-Correlation   & Self-generated    & Aperture image (2D)  \\
Self-holography with Antenna-Correlation  & Self-generated    & Measured correlations (1D)  \\ 
Cross-holography with Beam-Correlation  & Different antenna & Aperture image (2D)  \\ 
Cross-holography with Antenna-Correlation & Different antenna & Measured correlations (1D) \\ 
\hline
\end{tabular}
\end{table}

In this paper, we further develop the technique used in \citeA{2021RaSc...5607171K} to include self and cross-holography. We bring self and cross holographic techniques under the same framework and discuss their relative merits and demerits. We show that self-holography using Beam-Correlation can be performed with the visibilities alone. It is also shown that the two apparently distinct holographic approaches described in \citeA{8065418} and \citeA{2021RaSc...5607171K} are equivalent in their outcomes. As a demonstration of the techniques, we perform holography of the AAVS2. Both self and cross holographies are performed, and the results compared with \texttt{mfcal}. As this paper focuses on applying holography to calibrate the digitally beamformed SKA-Low stations, we will use the terms station, phased array or antenna array interchangeably to refer to these phased arrays. 

\section{Tensors for signal processing}
The theoretical framework developed in this paper makes use of multi-way data structures, also known as \textit{tensors} in the signal processing literature. While the expressions for holography can be written out without resorting to tensors, the use of tensors makes the equations tractable and provides novel insights into the technique. 


The following definitions are used in this paper. We would like point out that there are multiple (and often inconsistent) definitions of  tensors and associated data structures in the literature. 
\begin{enumerate}
    \item Arrays with any number of indices are referred to as tensors and denoted by calligraphic uppercase letters, such as $\mathcal{W}$. The total number of indices required to specify an element of a tensor is called its \textit{order}. \cite{2009SIAMR..51..455K}. Confusingly, the order of a tensor is also referred as its \textit{rank} sometimes, though the rank of a tensor can also denote a concept similar to matrix rank. 
    \item The elements constituting a tensor are denoted by the same letter as the tensor, but in plain lowercase with Greek alphabet indices, such as $w_{\alpha \beta}$. These indices can appear as subscript ($v_{\eta \tau}$), superscript ($v^{\eta \tau}$) or both ($v_{\tau}^{\eta}$). In this aspect, we deviate from the convention prevalent in signal processing where only subscripts are used \cite <e.g.,>{mimo_costa}, and adopt the tensor convention more common in physical sciences \cite <e.g.,>{1973grav.book.....M}
    \item Vectors are denoted by bold lowercase letters, such as $\bm{x}$. An element of a vector is denoted by a plain lowercase letter with a Greek alphabet index variable, such as $x_{\eta}$.
    \item Matrices are denoted by boldfaced uppercase letters such, as $\bm{W}$. Tensors of second-order can be represented by matrices.
\end{enumerate}

\subsection{Einstein summation}
We make use of Einstein summation for tensor operations. When Einstein summation is employed, identical indices appearing in \textit{both subscript and superscript} of the terms are summed over. Consequently, the index that is summed over is a \textit{dummy index} and may be replaced with a different letter. The indices of a tensor may need to be  \textit{raised} or \textit{lowered} for summation, which is achieved by contracting with the \textit{metric} tensor. The metric for the tensors used in this paper is Euclidean, with $\delta_{\eta \gamma}$ defining the metric. In other words, the metric can be represented by an identity matrix and therefore, raising or lowering of the indices is a trivial exercise. 

\section{Tensor formulation of phased array holography}
\label{sec:genholomath}
In this section we formulate holography using tensors which is applicable to both self and cross holographic techniques. We demonstrate that the distinct approaches to self-holography - namely the Beam-Correlation method and Antenna-Correlation method - yield the same results, although the former provides a 2D aperture image. We also show that self-holography can be carried out with visibilities alone.

We begin with formulating the Beam-Correlation holography method. In order to perform holography with phased arrays, the voltage beam of the station to be calibrated is scanned in a raster, and correlated with a reference beam pointed towards a strong source to obtain a sky image called the Beam-Correlation around the source. \change{This requires pointing the beam}{Scanning requires pointing the voltage beam of the station} towards each grid point on a pattern on the sky by a weighted addition of the individual antenna voltages. The phased array consists of $N$ individual antennas, each with an associated complex receive path gain $g_{\eta}; \eta \in [0, N-1]$. From each antenna \add{numbered $\eta$}, a stream of voltages \add{$v_{\tau}^{\eta}$} is acquired, with \add{$\tau$ being the time index and} $T$ being the total number of samples collected. Using Einstein summation, the collected voltages can be written as
\begin{eqnarray}
    \label{eq:volt_gain}
    v_{\tau}^{\eta} &= \delta^{\eta}_{\chi\phi} g^{\phi} s_{\tau}^{\chi}; \tau \in [0, T-1] \\
    \delta^{\eta}_{\chi\phi} &= \bigg\{
        \begin{array}{lr}
        1, \eta=\chi=\phi \\
        0, \rm{otherwise}
        \end{array}\label{eq:delta3index}
\end{eqnarray}
where the third-order tensor $\delta^{\eta}_{\chi\phi}$ defined by Eq.\ref{eq:delta3index} assists in performing Hadamard like multiplications\add{, with $\chi$ and $\phi$ as dummy indices.} \add{It may also be noted that in \protect{Eq.\ref{eq:volt_gain}}}\add{, we changed the dummy index of $g_{\eta}$ to $\phi$ and subsequently raised that index.} $s_{\tau}^{\chi}$ denote the voltages that would be collected if the system was ideal such that no \add{gain} calibration is required. \add{It also includes the noise added by the receiver chain, especially the noise contributions from the low noise amplifiers (LNAs) referenced to the system input. The receiver noise can also be explicitly included, however as we show in Sec.\protect{\ref{subsec:autoeffect}}}\add{, it manifests in autocorrelations alone and can safely be ignored, especially for cross-holography.}. Eq.\ref{eq:volt_gain} can also be written in linear algebra notation as $\bm{v_{\eta}} = g_{\eta} \bm{s_{\eta}}$ for each antenna, where $g_{\eta}$ is the gain of antenna $\eta$. \remove{It may also be noted that in \protect{Eq.\ref{eq:volt_gain}}}\remove{, we also changed the dummy index of $g_{\eta}$ and subsequently raised that index.}

The complex weights required to beamform the station to a grid of direction cosines $l,m$ on the sky can be organized into a third-order tensor $\mathcal{W}$ with elements $w_{\alpha \beta \eta}$. \add{We assume the station to be a planar array.} \change{For}{Then, for} a direction cosine pair $l_{\alpha}, m_{\beta}$ in the sky, the corresponding weight $w_{\alpha \beta \eta}$ to steer the voltage beam is
\begin{eqnarray}
\label{eq:weight_tensor}
    w_{\alpha \beta \eta} = exp[j2\pi\{x_{\eta}l_{\alpha} + y_{\eta}m_{\beta}\}],
\end{eqnarray}
where $x_{\eta}, y_{\eta}$ is the location of antenna $\eta$ in the aperture plane, in units of wavelengths. 
The third-order tensor of station voltage beams $\mathcal{D}$ over a raster pattern in the sky can then be written as,
\begin{eqnarray}
    \label{eq:V_einstein}
    d_{\alpha \beta \tau} &= w_{\alpha \beta \eta} \delta^{\eta}_{\chi\phi} z^{\phi} v_{\tau}^{\chi}.
\end{eqnarray}
where $\bm{z}$ is the steering vector of complex weights required to point the beam towards the calibrator. The multiplication of the antenna voltages with the steering vector removes the geometric phase for an off-boresight calibrator and brings it to the phase centre, ensuring that the Beam-Correlation is centred on the source. 

The Beam-Correlation $\bm{B}$ is obtained by cross-correlating the reference beam voltages with the scanned beams and temporal averaging. \add{For the cross-correlation to be meaningful, the reference beam voltages as well as the scanned beam voltages must be sampled at the same time, or interpolated to be aligned.} The vector consisting of the complex voltages from the reference antenna can be written as $\bm{r}$ and thus the generation of the Beam-Correlation becomes 
\begin{equation} 
\label{eq:beamco_mat}
    b_{\alpha \beta} = w_{\alpha \beta \eta} \delta^{\eta}_{\chi\phi} z^{\phi} v_{\tau}^{\chi} (r^{*})^{\tau},
\end{equation}
\add{where for brevity we have omitted a multiplicative factor of $1/T$, required for temporal averaging. This omission will result in an overall scaling factor requirement for absolute flux density calibration, however the relative gain measurements will be unaffected}. We now define a vector of "measured correlations" $\bm{h}$, such that 
\begin{equation}
\label{eq:vec_corr}
    h^{\eta} = \delta^{\eta}_{\chi\phi} z^{\phi} v_{\tau}^{\chi} (r^{*})^{\tau}
\end{equation}
is the cross-correlation between the antenna numbered $\eta$ and the reference beam. Then Eq.\ref{eq:beamco_mat} can be written compactly as 
\begin{equation}
    \label{eq:beamco_vec}
     b_{\alpha \beta} = w_{\alpha \beta \eta} h^{\eta}.
\end{equation}
Eq.\ref{eq:beamco_vec} describes a generalised form of holography, where the Beam-Correlation can be obtained from a vector of "measured correlations" $\bm{h}$. It is worth noting that these measured correlations $\bm{h}$ are used in the Antenna-Correlation holographic method to obtain the gains \cite{8065418}.

To obtain the complex aperture image $\bm{A}$, a Fourier transform of the Beam-Correlation $\bm{B}$ is taken. For this, we define a fourth-order "Fourier transform tensor" $\mathcal{F}$ such that
\begin{eqnarray}
\label{eq:ft_tensor}
    f_{\mu \sigma}^{\alpha \beta}  = exp[-j2\pi(x_{\mu}l_{\alpha}+y_{\sigma}m_{\beta})].
\end{eqnarray}
\add{where $\mu$ and $\sigma$ are the indices for the two orthogonal coordinates $x$ and $y$ in the aperture image plane.}
The aperture image can be written as
\begin{eqnarray}   
\label{eq:aper_image_1}
a_{\mu \sigma} 
   &= f_{\mu \sigma}^{\alpha \beta} b_{\alpha \beta} \nonumber \\ 
   &= f_{\mu \sigma}^{\alpha \beta} w_{\alpha \beta \eta} h^{\eta} .  
\end{eqnarray}
In Eq.\ref{eq:aper_image_1} the Fourier transform is \textit{independent} of the summation over the index $\eta$, and therefore can be treated separately for each antenna. In other words, each slice of the weight tensor $\mathcal{W}$ may be Fourier transformed separately. Thus, we can evaluate the Fourier transform of a slice of the weight tensor for each antenna $\eta$. 
In Eq.\ref{eq:fourier_expansion}, we combine the Fourier and beam-weight terms of Eq.\ref{eq:aper_image_1} by substituting Eqs.\ref{eq:weight_tensor} and \ref{eq:ft_tensor} and invoke the summation symbol explicitly for clarity, 
\begin{eqnarray}
\label{eq:fourier_expansion}
a_{\mu \sigma} 
&= \sum_{\eta} h_{\eta} \big( \sum_{\alpha} \sum_{\beta} exp[j2\pi \{ (x_{\eta}-x_{\mu})l_{\alpha}+(y_{\eta}-y_{\sigma})m_{\beta} \}] \big).
\end{eqnarray}
It is easy to observe that the operation in Eq.\ref{eq:fourier_expansion} is similar to a Fourier transform of sinusoids, resulting in delta functions in the Fourier plane. It can be seen that for a given $\eta$, when $\mu$ and $\sigma$ are such that $x_{\mu}=x_{\eta}$ and $y_{\sigma}=y_{\eta}$ corresponding to the location of the antenna $\eta$ in the aperture image, the exponential term becomes unity. As long as the direction cosines $l$ and $m$ span a large range, the double summation over the exponential approaches zero when $x_{\mu} \neq x_{\eta}$ and $y_{\sigma} \neq y_{\eta}$. \remove{Consequently, the resolution of the aperture image is limited by the extend of $l,m$ scan as per the uncertainty principle, as well as any windowing applied to the Beam-Correlation.} Thus, we may write
\begin{eqnarray}
\label{eq:aper_anloc_prop}
    a(x_{\mu} = x_{\eta}; y_{\sigma} = y_{\eta}) \propto h^{\eta}.
\end{eqnarray}
Substituting Eq.\ref{eq:volt_gain} in Eq.\ref{eq:vec_corr}, Eq.\ref{eq:aper_anloc_prop} becomes
\begin{eqnarray}
\label{eq:aper_anloc}
     a(x_{\mu} = x_{\eta}; y_{\sigma} = y_{\eta}) \propto \delta^{\eta}_{\chi\phi} z^{\phi} \delta^{\chi}_{\zeta\psi} g^{\psi} s_{\tau}^{\zeta} (r^{*})^{\tau}.
\end{eqnarray}
Eq.\ref{eq:aper_anloc} can then be rewritten in linear algebra as follows to enable an easier interpretation;
\begin{equation}
\label{eq:aper_anloc_la}
     a(x_{\mu} = x_{\eta}; y_{\sigma} = y_{\eta}) \propto
     z_{\eta} g_{\eta} \bm{s_{\eta}} \bm{r^*}, 
\end{equation}
where $z_{\eta} g_{\eta} \bm{s_{\eta}}$ is the $1 \times T$ \change{array}{row vector} of \add{complex} voltages from antenna $\eta$, including the corresponding complex gain $g_{\eta}$ and steering coefficient $z_{\eta}$, \change{$\bm{r}$ is the $1 \times T$ array of reference beam voltages}{$\bm{r}$ is the $T \times 1$ column vector of complex reference beam voltages}. Thus the interpretation of Eqs.\ref{eq:aper_anloc} and \ref{eq:aper_anloc_la} is that the aperture image obtained from holography consists of scaled versions of the complex correlation between the reference beam and individual antennas at the corresponding antenna locations. 

\remove{As the reference beam is common for all the antennas, the complex correlations at the antenna locations differ only by their relative receive path gains and variations in their embedded element patterns (EEPs) towards the calibrator. Under the assumption that the EEPs are all identical, we may extract relative receive path gains $k^{\eta}$ from the aperture image as} \add{We may now extract the relative receive path gains $k^{\eta}$ from the aperture image as} 
\begin{eqnarray}
\label{eq:aper_gains}
    k^{\eta} = a(x_{\mu} = x_{\eta}; y_{\sigma} = y_{\eta}) \propto h^{\eta}.
\end{eqnarray}
\add{As the reference beam is common for all the antennas, the complex measured correlations at the antenna locations differ only by their relative receive path gains and variations in their embedded element patterns (EEPs) towards the calibrator. If the EEPs are all identical, the gains obtained from holography will be proportional to the receive path such that $k^{\eta} \propto g^{\eta}$}. \change{This}{Also, the} relation between the aperture image $a_{\mu \sigma}$ at the antenna locations and the measured correlations $h^{\eta}$ show that the gains obtained from Beam-Correlation holography ought to be similar to those from the Antenna-Correlation method, except for multiplicative constants. Thus, the key takeaway messages from our analysis are as follows.
\begin{enumerate}
    \item The Beam-Correlation in Eq.\ref{eq:beamco_vec} is an inverse Fourier transform of the antenna locations weighted by the corresponding measured correlation\remove{, as the weight tensor given by Eq. \protect{\ref{eq:weight_tensor}}}.
    \item Consequently, a Fourier transform of the Beam-Correlation results in the measured correlations from the Antenna-Correlation method.
\end{enumerate}

\add{In writing \protect{Eqs.\ref{eq:fourier_expansion} to \ref{eq:aper_gains}}}\add{, an assumption is made that there are pixels in the aperture image corresponding to the exact antenna locations. In practice, the resolution of the aperture image is limited by the extent of $l,m$ scan. If the pixels do not correspond precisely to the antenna locations, scalloping will be seen in the aperture image. The effects of such leakage can be reduced with a combination of windowing and padding the Beam-Correlation to hyper-resolve the aperture image as done in Sec.\protect{\ref{sec:aavs2selfholo}}. As shown in the same section, the gains can be obtained from such an aperture image by taking an average of the pixels around an antenna locations, with the number of pixels to be averaged determined by the windowing used.}

\change{Therefore}{Thus,} either the measured correlations or the values from the antenna locations in the aperture image are useful for calibration, except for an overall scaling factor required to fix the flux density scale. A similar conclusion was reached in \citeA{9560415}, albeit with a different modelling. \add{The use of these gains is application dependent. For all-sky imaging with individual SKA-low stations, the gains can be used to calibrate the visibilities, for example by supplying them as \texttt{gains} file in a \texttt{miriad} dataset. On the other hand if the stations are used in beamformed mode, which is the expected use case for the SKA-low stations, the gains have to be used when the weights are calculated when the voltage beam is formed.}

\section{Self-holography}
\label{sec:selfholomath}
The reference beam in self-holography can be written as
\begin{eqnarray}
    r^{\tau} = z_{\gamma} v^{\gamma \tau}
\end{eqnarray}
where $\bm{z}$ is the vector of complex weights required to phase the beam towards the calibrator. This results in 
\begin{eqnarray}
\label{eq:corrmat}
     b_{\alpha \beta} &= w_{\alpha \beta \eta} \delta^{\eta}_{\chi\phi} z^{\phi} v_{\tau}^{\chi}  z^*_{\gamma} (v^*)^{\gamma \tau} \nonumber \\
      &= w_{\alpha \beta \eta}\delta^{\eta}_{\chi\phi} z^{\phi} v_{\tau}^{\chi} (v^*)^{\gamma \tau} z^*_{\gamma}  \nonumber \\
      &= w_{\alpha \beta \eta} \delta^{\eta}_{\chi\phi} z^{\phi} c^{\chi \gamma} z^*_{\gamma}
\end{eqnarray}
where we rearrange terms and identify $c^{\chi \gamma} = v^{\chi}_{\tau}(v^*)^{\gamma \tau}$ as the time averaged cross-correlation - visibility - between antennas $\chi$ and $\gamma$. The second-order tensor of visibilities can be represented as a correlation matrix $\bm{C}$. Comparing Eq.\ref{eq:corrmat} with Eq.\ref{eq:beamco_vec}, it can be observed that the measured correlations $\bm{h}$ can be obtained in self-holography from the correlation matrix,
\begin{equation}
\label{eq:meascorr_sh}
h^{\eta} = \delta^{\eta}_{\chi\phi} z^{\phi} c^{\chi \gamma} z^*_{\gamma}.   
\end{equation}
Eqs.\ref{eq:corrmat} and \ref{eq:meascorr_sh} show that the formation of the measured correlations, and consequently Beam-Correlation and aperture imaging, are all possible using visibilities alone for self-holography. While visibilities are computationally expensive compared to holography, the fact that they are useful for self-holography implies that direct comparisons can be made between various holography algorithms and conventional calibration techniques. 

\subsection{Effect of autocorrelations}
\label{subsec:autoeffect}
The correlation matrix in self-holography is Hermitian, with receiver noise power dominated autocorrelations occupying the diagonal. \add{If all the antennas have similar receiver noise, the autocorrelations will be similar across the antennas. However, for closely spaced antennas some amount of receiver noise gets coupled across antennas and can appear as an additional noise power in the cross-correlations, which is also a function of the antenna separation. Moreover the autocorrelations also contain the average component of the radio sky, particularly the diffuse emission from the Galaxy which is a function of the local sidereal time (LST).} It is therefore imperative that the effect of autocorrelations on the calibration results is understood. 

For this, we assume that the correlation matrix is diagonal, consisting of receiver noise powers alone \add{as autocorrelations}  and no signal originating in the sky. We write this as $c^{\chi \gamma} = 0, ~ \forall~\chi \neq \gamma$. We further assume that the steering is towards the boresight of the station, which implies $z^*_{\gamma} = 1$. This gives the measured correlations as $\bm{h} = diag(\bm{C})$. However, as already noted in Eq.\ref{eq:aper_gains}, $k^{\eta}\propto h^{\eta}$. Therefore in the general case, the gain obtained for each antenna will be biased by \change{its receiver noise power}{the excess power in its autocorrelations}, as self-holography is essentially a linear operation on the correlation matrix as shown in Eq.\ref{eq:meascorr_sh}. \remove{If all the antennas have similar receiver noise, the bias will be constant across the antennas. However, for closely spaced antennas some amount of receiver noise gets coupled across antennas and can appear as an additional noise power, which is also a function of the antenna separation. Moreover the autocorrelations also contain the average component of the radio sky, particularly the diffuse emission from the Galaxy which is a function of the local sidereal time (LST), further introducing LST dependent bias on the estimated gains. Nonetheless for an $N$ antenna array it can be seen that the contribution of the autocorrelation term is diminished by a factor of $N$ upon summation as per \protect{Eq.\ref{eq:meascorr_sh}}} \remove{Therefore, if the autocorrelations are not substantially higher than cross-correlations, the bias can be expected to be minimal. This was pointed out, in a qualitative sense, in \protect{\citeA{2021RaSc...5607171K}}}This bias could be substantial if the calibrator does not have sufficiently high flux densities so as to overcome the combined effects of the diffuse sky and the receiver noise. A potential solution to mitigate this bias is to split a station into two logical sub-arrays, form a reference beam with one of the sub-arrays and use it to calibrate the the other sub-array. Yet another potential solution within the paradigm of self-holography is an application of subspace filtering to extract data pertaining to the calibrator alone from the correlation matrix. However instead of these techniques, we propose cross-holography using a second phased array to provide the reference beam. This is discussed in \change{the next section}{Sec. \protect{\ref{sec:crossholo}}}.

\add{Nonetheless, for an $N$ antenna array the contribution of the autocorrelation terms to the gains is diminished by a factor of $N$ upon summation as per \protect{Eq.\ref{eq:meascorr_sh}}}. \add{Therefore, if a calibrator with high flux density such as the Sun is employed, autocorrelations will not be substantially higher than cross-correlations and the bias can be expected to be minimal. This was pointed out, in a qualitative sense, in \protect{\citeA{2021RaSc...5607171K}}. }

\section{Cross-holography}
\label{sec:crossholo}
An alternative to self-holography is using a reference beam from a \change{co-located, but physically distinct reference antenna}{nearby reference antenna, which is also far enough from the station under test to avoid cross-talk}. With cross-holography, the short intra-station baselines and autocorrelations do not contribute to the calibration. Using mathematics similar to the one in Sec.\ref{sec:selfholomath} this can be shown by elementary means if another phased array provides the reference beam. Therefore, while self-holography can potentially be biased by the excess power appearing in the autocorrelations and short intra-station baselines, cross-holography would be immune to receiver noise powers, and relatively insensitive to large-scale diffuse structures which resolve out on the longer inter-station baselines.

Similar to self-holography, cross-holography can be carried out with either Beam-Correlation or Antenna-Correlation. However, the inter-station visibilities between the antennas belonging to different stations are not available in general and therefore the correlation matrix approach is seldom a viable option for Beam-Correlation holography.   
\section{Holography applied to AAVS2}
In this section, we perform self and cross-holography of AAVS2. For cross-holography the reference beam is provided by the EDA2 station. The data were acquired on 30-Nov-2022, from 04:47 UTC onward. The Sun was chosen as the calibrator, and the observation frequency was centred on 159.375~MHz. Data were collected from a single polyphase filterbank (PFB) coarse channel of $\sim$ 925~kHz bandwidth \cite{2017JAI.....641015C}. The observations were carried out with the stations operating in different modes. AAVS2 - the station to be calibrated - was configured to perform voltage dumps of 262144 \add{complex} samples, corresponding to about 0.28s, from all the antennas every few seconds. For this work, we use one voltage dump dataset. EDA2, the station providing the reference beam, was configured to collect \add{the complex} station beam voltage data in \texttt{psrdada} format \cite{2021ascl.soft10003V}, with the beam pointing towards the Sun. As the two systems share a common clock source and have identical channelising schemes, the data from AAVS2 and EDA2 are coherent and suitable for cross-holography. \add{Additionally a \texttt{csv} file consisting of the locations of the antennas within AAVS2, obtained during the station deployment, is used to perform station voltage beam scan required for Beam-Correlation. The various holographic techniques are implemented in \texttt{python}. The tensor operations are carried out with the use of \texttt{numpy} package, specifically \textbf{numpy.einsum} is used for various operations involving large tensors.} We first perform self-holography with the AAVS2 data alone, followed by cross-holography including the EDA2 data. 
\subsection{Self-holography of AAVS2}
\label{sec:aavs2selfholo}
In order to validate the mathematics developed in Sec.\ref{sec:genholomath} and \ref{sec:selfholomath}, we perform self-holography with AAVS2 using the following three methods.
\begin{enumerate}
    \item Beam-Correlation from voltage dumps : We perform self-holography with voltage dumps using the same procedure outlined in \citeA{2021RaSc...5607171K} which is summarised here for completeness. The reference beam is formed from the voltages using the calculated position of the Sun over the observatory. The sky is gridded with dimensions 85 $\times$ 85, and the Beam-Correlation is calculated for this grid. A 2D Gaussian window of \change{$\sigma=0.5$}{$\sigma=1.0$ in direction cosine units giving a full width at half maximum (FWHM) of 2.355,} is applied to the Beam-Correlation to reduce ringing artefacts in the aperture image. \remove{and the} The Beam-Correlation is also zero padded to increase the final aperture image dimension to 513 $\times$ 513. Subsequently, the Beam-Correlation is Fourier transformed to obtain the complex aperture image. From the aperture image, \change{20}{10} pixels surrounding each antenna location are averaged to obtain the corresponding complex gains. 
    \item Beam-Correlation from visibilities : The voltage data for each antenna pair are correlated to obtain the correlation matrix consisting of visibilities. The Beam-Correlation is then obtained from this matrix, using Eq.\ref{eq:corrmat}, Fourier transformed and gains extracted as in the previous case.
    \item Cross-correlating the reference beam with individual antennas : From the voltage dumps, the measured correlations are calculated as per Eq.\ref{eq:vec_corr}, where each antenna is individually correlated with the reference beam. As the measured correlations are proportional to the receive path gains, they are normalised to be used as gains for calibration.
\end{enumerate}

\begin{figure}
\noindent\includegraphics[width=\textwidth]{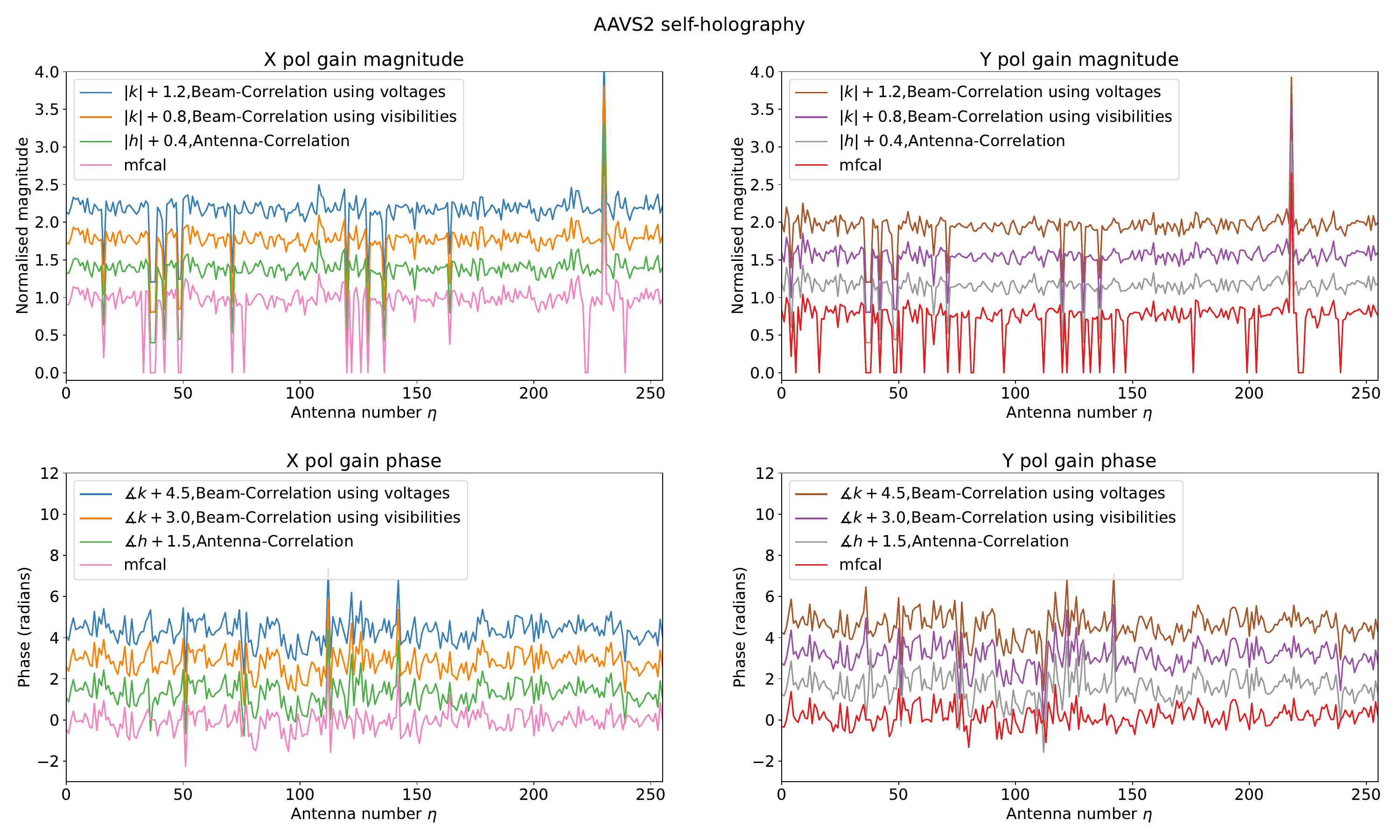}
\caption{Complex gains obtained from multiple self-holographic  techniques. Also shown are the gains as obtained from \texttt{mfcal} routine in \texttt{miriad}, for comparison. The \change{data}{holographic gains} are \remove{median filtered to set amplitude of outliers to zero and} normalised with antenna $\#3$ as the reference, similar to \texttt{mfcal}. The input visibility data to \texttt{mfcal} have several outlier antennas flagged. Therefore the corresponding \texttt{mfcal} gains are set to zero as seen in the plots; however these flags are not applicable for holography. In all plots the lines are artificially offset for clarity and the offset values are indicated in the legend. The gains from self-holographic techniques show extremely good agreement with each other as well as with the \texttt{mfcal} derived gains.}
\label{fig:sh_results}\note{Changed the plot labels to "Antenna-Correlation" instead of "measured correlations". Also expanded "BeamCorr" to "Beam-Correlation"}
\end{figure}

Fig.\ref{fig:sh_results} shows the results from self-holography using the above three methods, along with gains obtained from \texttt{mfcal} routine in \texttt{miriad}. The various self-holographic techniques yield the same results, validating the equivalence between them. The \texttt{mfcal} routine gives gains that are \add{broadly} consistent with the various self-holography methods, further increasing our confidence in the self-holography based calibration. \add{Pearson correlation coefficients calculated between various self-holographic techniques, as well as with cross-holography and \texttt{mfcal} are discussed in the \protect{Sec.\ref{sec:aavs2crossholo}}.}

\subsection{Cross-holography of AAVS2 with EDA2}
\label{sec:aavs2crossholo}
\begin{figure}
\noindent\includegraphics[width=\textwidth]{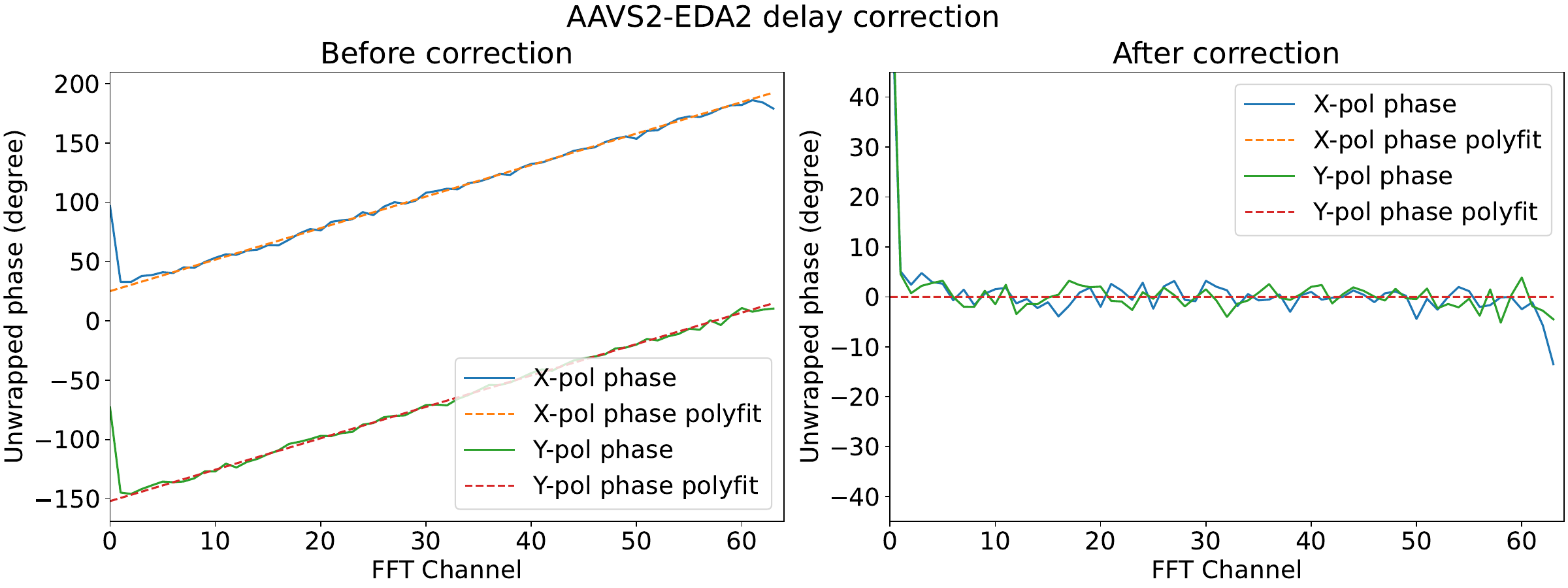}
\caption{Inter-station delay correction \add{for cross-holography}. The plots show the AAVS2-EDA2 cross-spectrum phase. The phase can be modelled by a delay and corrected for.}
\label{fig:delaycal}
\end{figure}

As AAVS2 and EDA2 are separated by $\sim$ 200m, a relative time delay is expected for a source that is away from the boresight \add{(zenith)}. Besides, the different lengths of the fibres carrying the RF data to the central facility can give rise to an additional delay. Prior to performing holography, this relative delay between the data from AAVS2 and EDA2 is estimated and corrected for. In order to coarse align the two different datasets, a finite number of samples are discarded from the EDA2 data using the difference between the timestamps in the corresponding observation metadata. The voltage data from AAVS2 are then beamformed towards the Sun, the resulting voltage beam channelised with FFT and cross-correlated with the similarly channelised EDA2 voltage stream. The number of FFT channels is a trade-off between sensitivity and bandwidth de-correlation effects. We deem 64 frequency channels across the $\sim$ 1~MHz bandwidth sufficient to reveal the relative delay without losing sensitivity. The phase of the resulting cross-spectrum clearly shows a linear ramp as a function of frequency, indicating a finite time delay between the stations. If the unwrapped phase gradient is more than $2\pi$ across the band, the data have to be aligned further. However, the timestamps in AAVS2 and EDA2 data are accurate enough that this is not required. A straight line model is fit to this phase ramp excluding the PFB edge channels, and used to re-phase the EDA2 data to obtain the reference beam. The phases and polynomial fits to them before and after the delay correction are shown in Fig.\ref{fig:delaycal}.

\begin{figure}
\noindent\includegraphics[width=\textwidth]{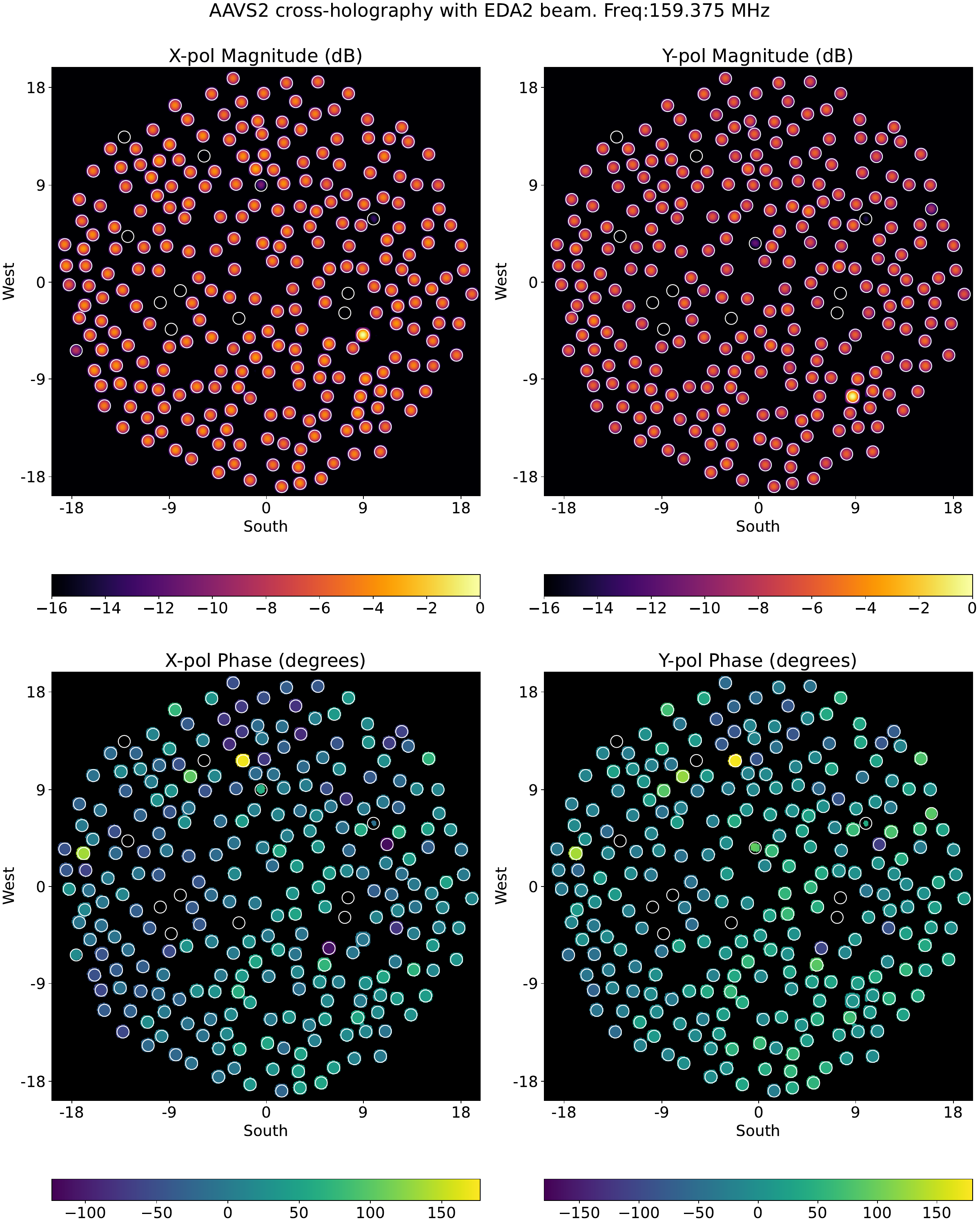}
\caption{Aperture images from cross-holography of AAVS2 with EDA2 reference beam. The magnitudes have been peak normalised and converted to decibels (dB) to enhance the dynamic range. All distances are in metres and the known antenna positions are circled. \add{Also, the phase images are masked using a mask created from the magnitude images.}}
\label{fig:ch_ap_image}
\end{figure}

\begin{figure}
\noindent\includegraphics[width=\textwidth]{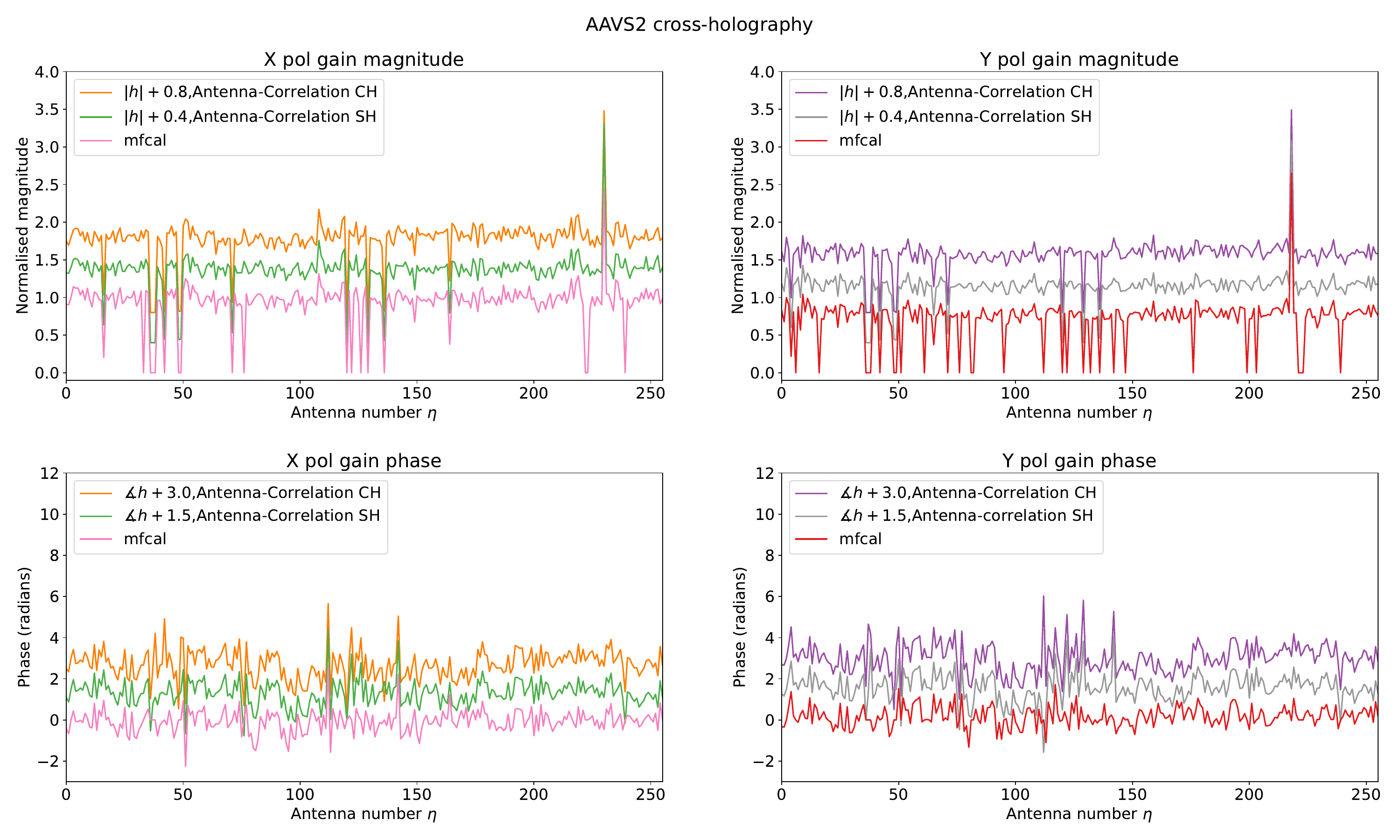}
\caption{Comparison between the complex gains obtained from cross-holography (CH) and self-holography (SH). \remove{The gains obtained from \texttt{mfcal} are also shown for comparison} \add{Shown are the measured correlations with the Antenna-Correlation approach along with the gains obtained from \texttt{mfcal} for comparison}. Similar to Fig.\ref{fig:sh_results} the flagged \add{\texttt{mfcal}} gains are set to zero and the lines are artificially offset for clarity.}\note{Changed the plot labels to "Antenna-Correlation" instead of "measured correlations"}
\label{fig:ch_results_plots}
\end{figure}

\begin{figure}
\noindent\includegraphics[width=1.1\textwidth]{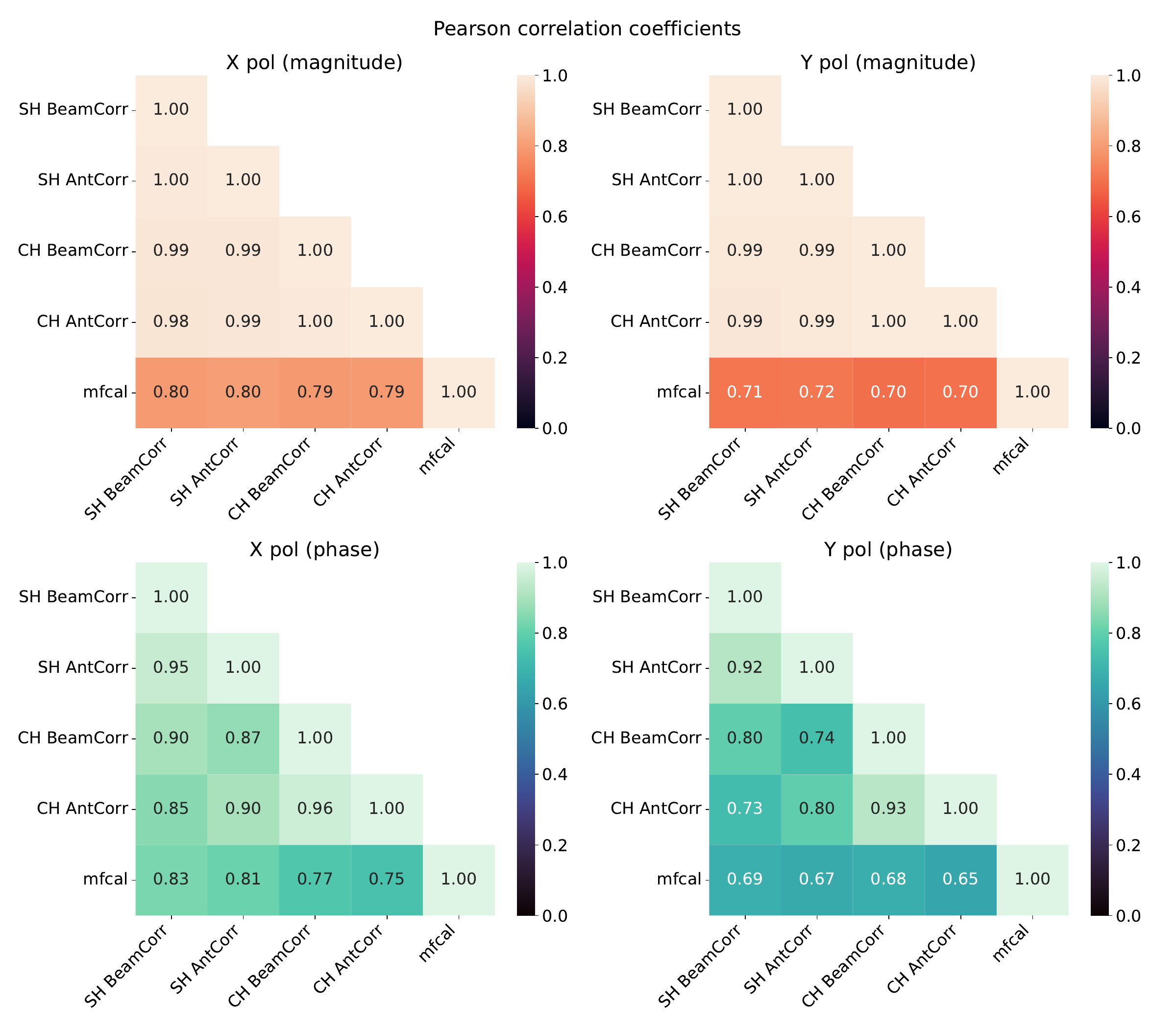}
\caption{Pearson correlation coefficients computed between the various gains shown in Figs. \ref{fig:sh_results} and \ref{fig:ch_results_plots}. A correlation coefficient of 1 denotes perfect correlation and 0 indicates lack of any correlation. The holographic techniques yield nearly identical gains, as shown by their high correlation. In general, the holographic gains are also consistent with \texttt{mfcal} gains. However they operate under different assumptions and use different algorithms, which is probably the reason for lower correlation coefficients between holographic methods and mfcal. The presence of flagged antennas in \texttt{mfcal} also contributes to a reduction in the correlation coefficients.}
\label{fig:pcorr_heatmaps}
\end{figure}

Subsequently, we perform cross-holography with both Beam-Correlation and Antenna-Correlation techniques. In Fig.\ref{fig:ch_ap_image}, the complex aperture images obtained from Beam-Correlation method are shown for both polarisations. Fig.\ref{fig:ch_results_plots} compares the measured correlations from \add{Antenna-Correlation} cross-holography with the ones from \add{Antenna-Correlation} self-holography and \texttt{mfcal}. \remove{The gains obtained from cross-holography show good agreement with self-holography as well as with the \texttt{mfcal} derived gains.} \add{In Fig.\protect{\ref{fig:pcorr_heatmaps}}}\add{,we show the Pearson correlation coefficient calculated between multiple calibration techniques, which we interpret as a measure of similarity between various techniques. Alternatively, the root-mean-square deviations (rmsd) between these quantities could also be employed, as Pearson correlation coefficient shows only a linear correlation. However rmsd is also quite sensitive to outliers. The various holographic techniques show very high correlation, which can be interpreted as them being consistent. An interesting observation is that holographic gains obtained from Beam-Correlation and Antenna-Correlation are very similar. Gains from self and cross-holographies show lower correlation, especially in their phases. Although the \texttt{mfcal} gains are correlated with holographic gains, they are not exactly the same. This is expected, as \texttt{mfcal} is fundamentally a different technique with different underlying assumptions, and several antennas had to be flagged for \texttt{mfcal} to converge. This is evident in the Y-polarisation \texttt{mfcal} gains in which several antennas are flagged.} 

\section{Discussion}
The results from Secs.\ref{sec:aavs2selfholo} and \ref{sec:aavs2crossholo} demonstrate the feasibility of applying multiple holographic techniques for calibrating the SKA-Low prototype stations. The convergence of various approaches towards identical outputs demonstrate the robustness of holographic techniques. Despite Beam-Correlation and Antenna-Correlation showing identical results, the two methods are not redundant as they serve different purposes. Owing to its simplicity, Antenna-Correlation technique can be used to obtain gains that can directly be used for station calibration. Beam-Correlation, on the other hand, provides a visual representation of the station gains where any systematics issues are easier to identify. As an example, a phase gradient would appear across the aperture image when a mis-calibration of the inter-station delay or source position causes an excess delay between two stations. \remove{The antennas in a station are grouped into "tiles" and any issue pertaining to specific tiles will be obvious in an aperture image.}\add{In a station, the 256 antennas are aggregated into groups of 16 at the SMART boxes, where the analog RF signals are also converted from electrical to optical. These optical signals are then patched to long distance fibre connection at the field node distribution hubs (FNDH) for transport to the control room, where they are connected to the tile processing modules (TPM, \protect{\citeA{2017JAI.....641014N}})}\add{with each TPM processing signals from 16 antennas \protect{\cite{2022JATIS...8a1010W}}}.
\add{Thus, owing to this grouping of antennas into "tiles" of 16 sharing some resources, an aperture image can be useful in identifying issues pertaining to specific SMART box or TPM signal paths.} Besides, the effect of any interfering source can be inferred by inspecting the Beam-Correlation, prior to any calibration and imaging. The effectiveness of the reference beam to isolate the calibrator from other sources in the sky can be verified with Beam-Correlation, as secondary sources in the sky that can potentially interfere with calibration (for example, via sidelobes) would be visible in the Beam-Correlation. This makes the Beam-Correlation method a powerful diagnostic tool.

\add{In general, the gains obtained from holography will be frequency and direction dependent. Therefore for broadband calibration of a station, holography (or any other calibration algorithm) has to be repeated at multiple frequencies so as to obtain the instrumental bandpass.}
Simulated EEPs of SKA-Low antennas in a station show considerable deviations from antenna to antenna, due to strong mutual coupling effects \cite{8879294, 2022JATIS...8a1023B}. The EEPs are distinct for each antenna, as well as each antenna EEP is a function of frequency, azimuth and elevation. While inclusion of EEPs into holography goes beyond the scope of this paper, a potential avenue to improve calibration is inclusion of simulated EEP of each antenna \add{to reduce the direction-dependence of the gains}. The tensor formalism introduced in this paper can be extended to include such multi-dimensional EEP data. If enough compact and bright sources to sufficiently sample the azimuth and elevation are identified, cross-holography utilising a reference antenna with accurately known beam patterns can be used to even verify the antenna EEPs.

\subsection{Choosing between self-holography and cross-holography}

The results from AAVS2 holography show that the derived gains are very similar for self and cross holography, implying that the impact of auto-correlations on self-holography is minimal for a bright source. Therefore, self-holography may suffice as a viable calibration technique for SKA-Low stations when Sun is available as a calibrator. Since multi-beam capabilities are built into SKA-Low station signal processing \cite{2016SPIE.9906E..2SH, ska_sysspec}, for observations demanding high accuracy one of the beams can be fixed on a calibrator and cross-correlated with each antenna within the station signal processing to track temporal gain changes in real-time. Such a near real-time station calibration is implemented in LOFAR \cite{2013A&A...556A...2V}, however it uses the correlation matrix and not holography.

However, self-holography relies on having a reference beam that can isolate the calibrator from other sources. The SKA-Low prototype station used in this work is close to being calibrated and therefore self-holography provides excellent results. On the other hand, if a station is significantly out of calibration (especially phase/delay calibration), the Beam-Correlation would be ill-formed and self-holography would fail. In such a scenario, cross-holography with a calibrated \add{reference} station will have to be employed\add{, such that the reference beam can isolate the calibrator.}

In the early phases of SKA-Low with a limited number of baselines, an iterative approach to holographic calibration is possible. A reference beam is formed with a partially calibrated station to bootstrap the overall process and used for holography of the other stations, and finally holography of the initial station is carried out with a reference beam from one of the other stations. \change{Alternatively, at each step the calibrated stations can be used to obtain a progressively narrower tied-array beam to calibrate rest of the stations}{Alternatively, once a station is calibrated, it can be combined with previously calibrated stations to obtain a progressively narrower tied-array reference beam to calibrate rest of the stations}. However, one potential issue to be considered for cross-station holography with the Sun is that at the frequencies of interest, the Sun gets resolved out on baselines longer than a few hundred metres. Therefore if cross-holography is to be used on such baselines, compact albeit weaker sources such as the A-team sources (Virgo-A, Hydra-A etc.) may be better suited. Nonetheless, cross-holography with the Sun will still be useful for a large number of SKA-Low stations which will be less than hundred metres apart, such as the closely spaced stations within the SKA-Low core or the ones within each cluster of 6 stations along the spiral arms \cite{ska_layout}. 
\add{\protect{\subsection{Can holography provide unknown antenna locations within an array ?}}}
\add{A question that frequently arises in phased array holography is whether it can be used to find unknown antenna locations in an array, especially using the Beam-Correlation technique which provides an aperture image. For phased arrays, the unknowns consist of both the antenna complex gains and the antenna locations forming the aperture. This is in contrast to dish holography where the unknown is the dish aperture alone. The dish beam scanning can be accomplished, by physically moving the aperture to be mapped. This does not require a-priori knowledge of the aperture, under the assumptions that that the aperture stays the same during the antenna movements and the dish pointing model is accurate. The unknown pixels in the dish aperture image can then be solved for with sufficient number of pointings. For phased arrays, even to scan the beam to obtain the Beam-Correlation, the antenna location information is required without which the steering vectors cannot be computed. 

If a sufficiently large bandwidth is used and the phase of an antenna across the band has a linear trend due to an excess delay, then it may be possible to to calculate its misalignment along the baseline to the reference station. However, the positions of the antennas within the array, the cable lengths from the antennas to the digitisers, and the phases introduced by the analog electronics all contribute to the gains obtained after holography. Therefore, such a delay could also arise from excess cable lengths. Nonetheless, accurate antenna positions within each station can be obtained during deployment so that small deviations like cable delays can be absorbed into the gains.}

\section{Conclusions}
In this work, we introduced a novel tensor based framework for phased array holography and unified multiple variants of holography, namely the Beam-Correlation and Antenna-Correlation techniques and their self and cross holography variants. The framework that we developed in this paper combined these different types of holography and shows that they are consistent among themselves and with standard interferometric calibration. The differences between self and cross holography, and the limitations related to applying self-holography to a fully uncalibrated station have also been discussed. The equations given in this paper can be extended to include multiple system non-idealities to improve calibration. The robustness of the techniques is demonstrated using on-sky data from AAVS2, and comparing the gains obtained with the ones from a more conventional interferometric calibration. The results are promising and in future, holography can potentially aid real-time station calibration of SKA-Low. 

\begin{acronyms}
    \acro{SKA-Low} Square Kilometre Array - Low
    \acro{AAVS2} Aperture Array Verification System 2
    \acro{EDA2} Engineering Development Array 2
    \acro{EEP} Embedded element pattern
    \acro{CH} Cross-holography
    \acro{SH} Self-holography
\end{acronyms}

%
\begin{notation}
\notation{$\mathcal{W}$} Third-order tensor of beamforming weights
\notation{$\mathcal{F}$} Fourth-order tensor for Fourier tranform
\notation{$\mathcal{D}$} third-order tensor of station voltage beams over the sky
\notation{$\bm{A}$} Complex aperture image
\notation{$\bm{B}$} Beam-Correlation
\notation{$\bm{C}$} Correlation matrix
\notation{$x,y$} Location of antennas in the aperture plane in units of wavelength
\notation{$l,m$} Direction cosines in the sky, with respect to the array phase centre
\notation{$\bm{h}$} Measured correlations
\notation{$\bm{g}$} Receiver gains
\end{notation}

\section{Open Research}
Code and data used for this paper are available in a dataset associated with this publication on Zenodo \cite{zenodo_upload}.

\acknowledgments
This work makes use of Inyarrimanha Ilgari Bundara (Murchison Radio-astronomy Observatory), operated by CSIRO. We acknowledge the Wajarri Yamatji people as the traditional owners of the Observatory site. AAVS2 and EDA2 are hosted by the Murchison Widefield Array under an agreement via the MWA External Instruments Policy. We acknowledge the efforts of INAF in building the AAVS2 and providing TPMs for EDA2. The acquisition systems for AAVS2 and EDA2 were designed and purchased by INAF/Oxford University and the receiver chain was design by INAF, as part of the SKA design and prototyping program. We also acknowledge the ICRAR-Curtin operations team for maintaining the AAVS2 and EDA2 systems. This works also makes use of the NASA Astrophysics Data System (ADS). This work also uses the following python packages and we would like to thank the authors and maintainers of these packages - \texttt{numpy} \cite{2020Natur.585..357H}, \texttt{astropy} \cite{astropy:2013, astropy:2018}, \texttt{matplotlib} \cite{Hunter:2007}, \texttt{seaborn} \cite{Waskom2021}, \texttt{pandas} \cite{pandas1, pandas2} and \texttt{h5py} (\url{https://www.h5py.org/}). \add{We also thank the reviewers for their insightful comments which have helped us to improve this paper.}


%
%



\bibliography{AAVS_holography}

\begin{thebibliography}{}

\bibitem [\protect \citeauthoryear {%
{Astropy Collaboration}%
\ \protect \BOthers {.}}{%
{Astropy Collaboration}%
\ \protect \BOthers {.}}{%
{\protect \APACyear {2018}}%
}]{%
astropy:2018}
\APACinsertmetastar {%
astropy:2018}%
\begin{APACrefauthors}%
{Astropy Collaboration}%
, {Price-Whelan}, A\BPBI M.%
, {Sip{\H{o}}cz}, B\BPBI M.%
, {G{\"u}nther}, H\BPBI M.%
, {Lim}, P\BPBI L.%
, {Crawford}, S\BPBI M.%
\BDBL {}{Astropy Contributors}%
\end{APACrefauthors}%
\unskip\
\newblock
\APACrefYearMonthDay{2018}{{\APACmonth{09}}}{}.
\newblock
{\BBOQ}\APACrefatitle {{The Astropy Project: Building an Open-science Project and Status of the v2.0 Core Package}} {{The Astropy Project: Building an Open-science Project and Status of the v2.0 Core Package}}.{\BBCQ}
\newblock
\APACjournalVolNumPages{The Astronomical Journal}{156}{3}{123}.
\newblock
\begin{APACrefDOI} \doi{10.3847/1538-3881/aabc4f} \end{APACrefDOI}
\PrintBackRefs{\CurrentBib}

\bibitem [\protect \citeauthoryear {%
{Astropy Collaboration}%
\ \protect \BOthers {.}}{%
{Astropy Collaboration}%
\ \protect \BOthers {.}}{%
{\protect \APACyear {2013}}%
}]{%
astropy:2013}
\APACinsertmetastar {%
astropy:2013}%
\begin{APACrefauthors}%
{Astropy Collaboration}%
, {Robitaille}, T\BPBI P.%
, {Tollerud}, E\BPBI J.%
, {Greenfield}, P.%
, {Droettboom}, M.%
, {Bray}, E.%
\BDBL {}{Streicher}, O.%
\end{APACrefauthors}%
\unskip\
\newblock
\APACrefYearMonthDay{2013}{{\APACmonth{10}}}{}.
\newblock
{\BBOQ}\APACrefatitle {{Astropy: A community Python package for astronomy}} {{Astropy: A community Python package for astronomy}}.{\BBCQ}
\newblock
\APACjournalVolNumPages{Astronomy \& Astrophysics}{558}{}{A33}.
\newblock
\begin{APACrefDOI} \doi{10.1051/0004-6361/201322068} \end{APACrefDOI}
\PrintBackRefs{\CurrentBib}

\bibitem [\protect \citeauthoryear {%
{Benthem}%
\ \protect \BOthers {.}}{%
{Benthem}%
\ \protect \BOthers {.}}{%
{\protect \APACyear {2021}}%
}]{%
2021A&A...655A...5B}
\APACinsertmetastar {%
2021A&A...655A...5B}%
\begin{APACrefauthors}%
{Benthem}, P.%
, {Wayth}, R.%
, {de Lera Acedo}, E.%
, {Zarb Adami}, K.%
, {Alderighi}, M.%
, {Belli}, C.%
\BDBL {}{Williams}, A.%
\end{APACrefauthors}%
\unskip\
\newblock
\APACrefYearMonthDay{2021}{{\APACmonth{11}}}{}.
\newblock
{\BBOQ}\APACrefatitle {{The Aperture Array Verification System 1: System overview and early commissioning results}} {{The Aperture Array Verification System 1: System overview and early commissioning results}}.{\BBCQ}
\newblock
\APACjournalVolNumPages{Astronomy \& Astrophysics}{655}{}{A5}.
\newblock
\begin{APACrefDOI} \doi{10.1051/0004-6361/202040086} \end{APACrefDOI}
\PrintBackRefs{\CurrentBib}

\bibitem [\protect \citeauthoryear {%
{Bolli}%
, {Bercigli}%
, {Di Ninni}%
, {Mezzadrelli}%
\BCBL {}\ \BBA {} {Virone}%
}{%
{Bolli}%
\ \protect \BOthers {.}}{%
{\protect \APACyear {2022}}%
}]{%
2022JATIS...8a1023B}
\APACinsertmetastar {%
2022JATIS...8a1023B}%
\begin{APACrefauthors}%
{Bolli}, P.%
, {Bercigli}, M.%
, {Di Ninni}, P.%
, {Mezzadrelli}, L.%
\BCBL {}\ \BBA {} {Virone}, G.%
\end{APACrefauthors}%
\unskip\
\newblock
\APACrefYearMonthDay{2022}{{\APACmonth{01}}}{}.
\newblock
{\BBOQ}\APACrefatitle {{Impact of mutual coupling between SKALA4.1 antennas to the spectral smoothness response}} {{Impact of mutual coupling between SKALA4.1 antennas to the spectral smoothness response}}.{\BBCQ}
\newblock
\APACjournalVolNumPages{Journal of Astronomical Telescopes, Instruments, and Systems}{8}{}{011023}.
\newblock
\begin{APACrefDOI} \doi{10.1117/1.JATIS.8.1.011023} \end{APACrefDOI}
\PrintBackRefs{\CurrentBib}

\bibitem [\protect \citeauthoryear {%
Bolli%
\ \protect \BOthers {.}}{%
Bolli%
\ \protect \BOthers {.}}{%
{\protect \APACyear {2020}}%
}]{%
9107113}
\APACinsertmetastar {%
9107113}%
\begin{APACrefauthors}%
Bolli, P.%
, Mezzadrelli, L.%
, Monari, J.%
, Perini, F.%
, Tibaldi, A.%
, Virone, G.%
\BDBL {}Schiaffino, M.%
\end{APACrefauthors}%
\unskip\
\newblock
\APACrefYearMonthDay{2020}{}{}.
\newblock
{\BBOQ}\APACrefatitle {Test-Driven Design of an Active Dual-Polarized Log-Periodic Antenna for the Square Kilometre Array} {Test-driven design of an active dual-polarized log-periodic antenna for the square kilometre array}.{\BBCQ}
\newblock
\APACjournalVolNumPages{IEEE Open Journal of Antennas and Propagation}{1}{}{253-263}.
\newblock
\begin{APACrefDOI} \doi{10.1109/OJAP.2020.2999109} \end{APACrefDOI}
\PrintBackRefs{\CurrentBib}

\bibitem [\protect \citeauthoryear {%
{Caiazzo}%
}{%
{Caiazzo}%
}{%
{\protect \APACyear {2017}}%
}]{%
ska_sysspec}
\APACinsertmetastar {%
ska_sysspec}%
\begin{APACrefauthors}%
{Caiazzo}, M.%
\end{APACrefauthors}%
\unskip\
\newblock
\APACrefYearMonthDay{2017}{}{}.
\newblock
\APACrefbtitle {{SKA Phase 1 System Requirements Specification}} {{SKA Phase 1 System Requirements Specification}}\ \APACbVolEdTR{}{\BTR{}\ \BNUM\ SKA-TEL-SKO-0000008}.
\newblock
\APACaddressInstitution{}{SKAO}.
\newblock
\begin{APACrefURL} \url{https://www.skao.int/sites/default/files/documents/d3-SKA-TEL-SKO-0000008-Rev11_SKA1SystemRequirementSpecification.pdf} \end{APACrefURL}
\PrintBackRefs{\CurrentBib}

\bibitem [\protect \citeauthoryear {%
{Comoretto}%
\ \protect \BOthers {.}}{%
{Comoretto}%
\ \protect \BOthers {.}}{%
{\protect \APACyear {2017}}%
}]{%
2017JAI.....641015C}
\APACinsertmetastar {%
2017JAI.....641015C}%
\begin{APACrefauthors}%
{Comoretto}, G.%
, {Chiello}, R.%
, {Roberts}, M.%
, {Halsall}, R.%
, {Adami}, K\BPBI Z.%
, {Alderighi}, M.%
\BDBL {}{Zaccaro}, E.%
\end{APACrefauthors}%
\unskip\
\newblock
\APACrefYearMonthDay{2017}{{\APACmonth{03}}}{}.
\newblock
{\BBOQ}\APACrefatitle {{The Signal Processing Firmware for the Low Frequency Aperture Array}} {{The Signal Processing Firmware for the Low Frequency Aperture Array}}.{\BBCQ}
\newblock
\APACjournalVolNumPages{Journal of Astronomical Instrumentation}{6}{1}{1641015}.
\newblock
\begin{APACrefDOI} \doi{10.1142/S2251171716410154} \end{APACrefDOI}
\PrintBackRefs{\CurrentBib}

\bibitem [\protect \citeauthoryear {%
{Costa}%
\ \BBA {} {Haykin}%
}{%
{Costa}%
\ \BBA {} {Haykin}%
}{%
{\protect \APACyear {2010}}%
}]{%
mimo_costa}
\APACinsertmetastar {%
mimo_costa}%
\begin{APACrefauthors}%
{Costa}, N.%
\BCBT {}\ \BBA {} {Haykin}, S.%
\end{APACrefauthors}%
\unskip\
\newblock
\APACrefYear{2010}.
\newblock
\APACrefbtitle {{Multiple‐Input, Multiple‐Output Channel Models}} {{Multiple‐Input, Multiple‐Output Channel Models}}.
\newblock
\begin{APACrefDOI} \doi{10.1002/9780470590676} \end{APACrefDOI}
\PrintBackRefs{\CurrentBib}

\bibitem [\protect \citeauthoryear {%
Davidson%
\ \protect \BOthers {.}}{%
Davidson%
\ \protect \BOthers {.}}{%
{\protect \APACyear {2019}}%
}]{%
8879294}
\APACinsertmetastar {%
8879294}%
\begin{APACrefauthors}%
Davidson, D\BPBI B.%
, Bolli, P.%
, Bercigli, M.%
, di Ninni, P.%
, Steiner, R.%
, Tingay, S.%
\BDBL {}Wayth, R.%
\end{APACrefauthors}%
\unskip\
\newblock
\APACrefYearMonthDay{2019}{}{}.
\newblock
{\BBOQ}\APACrefatitle {Electromagnetic modelling of the SKA-LOW AAVS1.5 prototype} {Electromagnetic modelling of the ska-low aavs1.5 prototype}.{\BBCQ}
\newblock
\BIn{} \APACrefbtitle {2019 International Conference on Electromagnetics in Advanced Applications (ICEAA)} {2019 international conference on electromagnetics in advanced applications (iceaa)}\ (\BPG~1032-1037).
\newblock
\begin{APACrefDOI} \doi{10.1109/ICEAA.2019.8879294} \end{APACrefDOI}
\PrintBackRefs{\CurrentBib}

\bibitem [\protect \citeauthoryear {%
{Dewdney}%
\ \BBA {} {Braun}%
}{%
{Dewdney}%
\ \BBA {} {Braun}%
}{%
{\protect \APACyear {2016}}%
}]{%
ska_layout}
\APACinsertmetastar {%
ska_layout}%
\begin{APACrefauthors}%
{Dewdney}, P\BPBI E.%
\BCBT {}\ \BBA {} {Braun}, R.%
\end{APACrefauthors}%
\unskip\
\newblock
\APACrefYearMonthDay{2016}{}{}.
\newblock
\APACrefbtitle {{SKA1-LOW CONFIGURATION COORDINATES – COMPLETE SET}} {{SKA1-LOW CONFIGURATION COORDINATES – COMPLETE SET}}\ \APACbVolEdTR{}{\BTR{}\ \BNUM\ SKA-TEL-SKO-0000422}.
\newblock
\APACaddressInstitution{}{SKAO}.
\newblock
\begin{APACrefURL} \url{https://www.skao.int/sites/default/files/documents/d18-SKA-TEL-SKO-0000422_02_SKA1_LowConfigurationCoordinates-1.pdf} \end{APACrefURL}
\PrintBackRefs{\CurrentBib}

\bibitem [\protect \citeauthoryear {%
{Hampson}%
, {Bunton}%
, {Gunst}%
, {Baillie}%
\BCBL {}\ \BBA {} {bij de Vaate}%
}{%
{Hampson}%
\ \protect \BOthers {.}}{%
{\protect \APACyear {2016}}%
}]{%
2016SPIE.9906E..2SH}
\APACinsertmetastar {%
2016SPIE.9906E..2SH}%
\begin{APACrefauthors}%
{Hampson}, G\BPBI A.%
, {Bunton}, J\BPBI D.%
, {Gunst}, A\BPBI W.%
, {Baillie}, P.%
\BCBL {}\ \BBA {} {bij de Vaate}, J\BHBI G.%
\end{APACrefauthors}%
\unskip\
\newblock
\APACrefYearMonthDay{2016}{{\APACmonth{07}}}{}.
\newblock
{\BBOQ}\APACrefatitle {{Introduction to the SKA low correlator and beamformer system}} {{Introduction to the SKA low correlator and beamformer system}}.{\BBCQ}
\newblock
\BIn{} H\BPBI J.~{Hall}, R.~{Gilmozzi}\BCBL {}\ \BBA {} H\BPBI K.~{Marshall}\ (\BEDS), \APACrefbtitle {Ground-based and Airborne Telescopes VI} {Ground-based and airborne telescopes vi}\ (\BVOL\ 9906, \BPG~99062S).
\newblock
\begin{APACrefDOI} \doi{10.1117/12.2231524} \end{APACrefDOI}
\PrintBackRefs{\CurrentBib}

\bibitem [\protect \citeauthoryear {%
{Harris}%
\ \protect \BOthers {.}}{%
{Harris}%
\ \protect \BOthers {.}}{%
{\protect \APACyear {2020}}%
}]{%
2020Natur.585..357H}
\APACinsertmetastar {%
2020Natur.585..357H}%
\begin{APACrefauthors}%
{Harris}, C\BPBI R.%
, {Millman}, K\BPBI J.%
, {van der Walt}, S\BPBI J.%
, {Gommers}, R.%
, {Virtanen}, P.%
, {Cournapeau}, D.%
\BDBL {}{Oliphant}, T\BPBI E.%
\end{APACrefauthors}%
\unskip\
\newblock
\APACrefYearMonthDay{2020}{{\APACmonth{09}}}{}.
\newblock
{\BBOQ}\APACrefatitle {{Array programming with NumPy}} {{Array programming with NumPy}}.{\BBCQ}
\newblock
\APACjournalVolNumPages{Nature}{585}{7825}{357-362}.
\newblock
\begin{APACrefDOI} \doi{10.1038/s41586-020-2649-2} \end{APACrefDOI}
\PrintBackRefs{\CurrentBib}

\bibitem [\protect \citeauthoryear {%
Hunter%
}{%
Hunter%
}{%
{\protect \APACyear {2007}}%
}]{%
Hunter:2007}
\APACinsertmetastar {%
Hunter:2007}%
\begin{APACrefauthors}%
Hunter, J\BPBI D.%
\end{APACrefauthors}%
\unskip\
\newblock
\APACrefYearMonthDay{2007}{}{}.
\newblock
{\BBOQ}\APACrefatitle {Matplotlib: A 2D graphics environment} {Matplotlib: A 2d graphics environment}.{\BBCQ}
\newblock
\APACjournalVolNumPages{Computing in Science \& Engineering}{9}{3}{90--95}.
\newblock
\begin{APACrefDOI} \doi{10.1109/MCSE.2007.55} \end{APACrefDOI}
\PrintBackRefs{\CurrentBib}

\bibitem [\protect \citeauthoryear {%
{Kiefner}%
, {Wayth}%
, {Davidson}%
\BCBL {}\ \BBA {} {Sokolowski}%
}{%
{Kiefner}%
\ \protect \BOthers {.}}{%
{\protect \APACyear {2021}}%
}]{%
2021RaSc...5607171K}
\APACinsertmetastar {%
2021RaSc...5607171K}%
\begin{APACrefauthors}%
{Kiefner}, U.%
, {Wayth}, R\BPBI B.%
, {Davidson}, D\BPBI B.%
\BCBL {}\ \BBA {} {Sokolowski}, M.%
\end{APACrefauthors}%
\unskip\
\newblock
\APACrefYearMonthDay{2021}{{\APACmonth{05}}}{}.
\newblock
{\BBOQ}\APACrefatitle {{Holographic Calibration of Phased Array Telescopes}} {{Holographic Calibration of Phased Array Telescopes}}.{\BBCQ}
\newblock
\APACjournalVolNumPages{Radio Science}{56}{5}{e07171}.
\newblock
\begin{APACrefDOI} \doi{10.1029/2020RS007171} \end{APACrefDOI}
\PrintBackRefs{\CurrentBib}

\bibitem [\protect \citeauthoryear {%
{Kolda}%
\ \BBA {} {Bader}%
}{%
{Kolda}%
\ \BBA {} {Bader}%
}{%
{\protect \APACyear {2009}}%
}]{%
2009SIAMR..51..455K}
\APACinsertmetastar {%
2009SIAMR..51..455K}%
\begin{APACrefauthors}%
{Kolda}, T\BPBI G.%
\BCBT {}\ \BBA {} {Bader}, B\BPBI W.%
\end{APACrefauthors}%
\unskip\
\newblock
\APACrefYearMonthDay{2009}{{\APACmonth{01}}}{}.
\newblock
{\BBOQ}\APACrefatitle {{Tensor Decompositions and Applications}} {{Tensor Decompositions and Applications}}.{\BBCQ}
\newblock
\APACjournalVolNumPages{SIAM Review}{51}{3}{455-500}.
\newblock
\begin{APACrefDOI} \doi{10.1137/07070111X} \end{APACrefDOI}
\PrintBackRefs{\CurrentBib}

\bibitem [\protect \citeauthoryear {%
{Macario}%
\ \protect \BOthers {.}}{%
{Macario}%
\ \protect \BOthers {.}}{%
{\protect \APACyear {2022}}%
}]{%
2022JATIS...8a1014M}
\APACinsertmetastar {%
2022JATIS...8a1014M}%
\begin{APACrefauthors}%
{Macario}, G.%
, {Pupillo}, G.%
, {Bernardi}, G.%
, {Bolli}, P.%
, {Di Ninni}, P.%
, {Comoretto}, G.%
\BDBL {}{Bhushan}, R.%
\end{APACrefauthors}%
\unskip\
\newblock
\APACrefYearMonthDay{2022}{{\APACmonth{01}}}{}.
\newblock
{\BBOQ}\APACrefatitle {{Characterization of the SKA1-Low prototype station Aperture Array Verification System 2}} {{Characterization of the SKA1-Low prototype station Aperture Array Verification System 2}}.{\BBCQ}
\newblock
\APACjournalVolNumPages{Journal of Astronomical Telescopes, Instruments, and Systems}{8}{}{011014}.
\newblock
\begin{APACrefDOI} \doi{10.1117/1.JATIS.8.1.011014} \end{APACrefDOI}
\PrintBackRefs{\CurrentBib}

\bibitem [\protect \citeauthoryear {%
{Misner}%
, {Thorne}%
\BCBL {}\ \BBA {} {Wheeler}%
}{%
{Misner}%
\ \protect \BOthers {.}}{%
{\protect \APACyear {1973}}%
}]{%
1973grav.book.....M}
\APACinsertmetastar {%
1973grav.book.....M}%
\begin{APACrefauthors}%
{Misner}, C\BPBI W.%
, {Thorne}, K\BPBI S.%
\BCBL {}\ \BBA {} {Wheeler}, J\BPBI A.%
\end{APACrefauthors}%
\unskip\
\newblock
\APACrefYear{1973}.
\newblock
\APACrefbtitle {{Gravitation}} {{Gravitation}}.
\PrintBackRefs{\CurrentBib}

\bibitem [\protect \citeauthoryear {%
{Naldi}%
\ \protect \BOthers {.}}{%
{Naldi}%
\ \protect \BOthers {.}}{%
{\protect \APACyear {2017}}%
}]{%
2017JAI.....641014N}
\APACinsertmetastar {%
2017JAI.....641014N}%
\begin{APACrefauthors}%
{Naldi}, G.%
, {Mattana}, A.%
, {Pastore}, S.%
, {Alderighi}, M.%
, {Zarb Adami}, K.%
, {Schillir{\`o}}, F.%
\BDBL {}{Zaccaro}, E.%
\end{APACrefauthors}%
\unskip\
\newblock
\APACrefYearMonthDay{2017}{{\APACmonth{03}}}{}.
\newblock
{\BBOQ}\APACrefatitle {{The Digital Signal Processing Platform for the Low Frequency Aperture Array: Preliminary Results on the Data Acquisition Unit}} {{The Digital Signal Processing Platform for the Low Frequency Aperture Array: Preliminary Results on the Data Acquisition Unit}}.{\BBCQ}
\newblock
\APACjournalVolNumPages{Journal of Astronomical Instrumentation}{6}{1}{1641014}.
\newblock
\begin{APACrefDOI} \doi{10.1142/S2251171716410142} \end{APACrefDOI}
\PrintBackRefs{\CurrentBib}

\bibitem [\protect \citeauthoryear {%
{Salas}%
, {Brentjens}%
, {Bordenave}%
, {Oonk}%
\BCBL {}\ \BBA {} {R{\"o}ttgering}%
}{%
{Salas}%
\ \protect \BOthers {.}}{%
{\protect \APACyear {2020}}%
}]{%
2020A&A...635A.207S}
\APACinsertmetastar {%
2020A&A...635A.207S}%
\begin{APACrefauthors}%
{Salas}, P.%
, {Brentjens}, M\BPBI A.%
, {Bordenave}, D\BPBI D.%
, {Oonk}, J\BPBI B\BPBI R.%
\BCBL {}\ \BBA {} {R{\"o}ttgering}, H\BPBI J\BPBI A.%
\end{APACrefauthors}%
\unskip\
\newblock
\APACrefYearMonthDay{2020}{{\APACmonth{03}}}{}.
\newblock
{\BBOQ}\APACrefatitle {{Tied-array holography with LOFAR}} {{Tied-array holography with LOFAR}}.{\BBCQ}
\newblock
\APACjournalVolNumPages{Astronomy \& Astrophysics}{635}{}{A207}.
\newblock
\begin{APACrefDOI} \doi{10.1051/0004-6361/201935670} \end{APACrefDOI}
\PrintBackRefs{\CurrentBib}

\bibitem [\protect \citeauthoryear {%
{Sault}%
, {Teuben}%
\BCBL {}\ \BBA {} {Wright}%
}{%
{Sault}%
\ \protect \BOthers {.}}{%
{\protect \APACyear {1995}}%
}]{%
1995ASPC...77..433S}
\APACinsertmetastar {%
1995ASPC...77..433S}%
\begin{APACrefauthors}%
{Sault}, R\BPBI J.%
, {Teuben}, P\BPBI J.%
\BCBL {}\ \BBA {} {Wright}, M\BPBI C\BPBI H.%
\end{APACrefauthors}%
\unskip\
\newblock
\APACrefYearMonthDay{1995}{{\APACmonth{01}}}{}.
\newblock
{\BBOQ}\APACrefatitle {{A Retrospective View of MIRIAD}} {{A Retrospective View of MIRIAD}}.{\BBCQ}
\newblock
\BIn{} R\BPBI A.~{Shaw}, H\BPBI E.~{Payne}\BCBL {}\ \BBA {} J\BPBI J\BPBI E.~{Hayes}\ (\BEDS), \APACrefbtitle {Astronomical Data Analysis Software and Systems IV} {Astronomical data analysis software and systems iv}\ (\BVOL~77, \BPG~433).
\PrintBackRefs{\CurrentBib}

\bibitem [\protect \citeauthoryear {%
{Scott}%
\ \BBA {} {Ryle}%
}{%
{Scott}%
\ \BBA {} {Ryle}%
}{%
{\protect \APACyear {1977}}%
}]{%
1977MNRAS.178..539S}
\APACinsertmetastar {%
1977MNRAS.178..539S}%
\begin{APACrefauthors}%
{Scott}, P\BPBI F.%
\BCBT {}\ \BBA {} {Ryle}, M.%
\end{APACrefauthors}%
\unskip\
\newblock
\APACrefYearMonthDay{1977}{{\APACmonth{03}}}{}.
\newblock
{\BBOQ}\APACrefatitle {{A rapid method for measuring the figure of a radio telescope reflector.}} {{A rapid method for measuring the figure of a radio telescope reflector.}}{\BBCQ}
\newblock
\APACjournalVolNumPages{Monthly Notices of the Royal Astronomical Society}{178}{}{539-545}.
\newblock
\begin{APACrefDOI} \doi{10.1093/mnras/178.4.539} \end{APACrefDOI}
\PrintBackRefs{\CurrentBib}

\bibitem [\protect \citeauthoryear {%
{Sokolowski}%
\ \protect \BOthers {.}}{%
{Sokolowski}%
\ \protect \BOthers {.}}{%
{\protect \APACyear {2021}}%
}]{%
2021PASA...38...23S}
\APACinsertmetastar {%
2021PASA...38...23S}%
\begin{APACrefauthors}%
{Sokolowski}, M.%
, {Wayth}, R\BPBI B.%
, {Bhat}, N\BPBI D\BPBI R.%
, {Price}, D.%
, {Broderick}, J\BPBI W.%
, {Bernardi}, G.%
\BDBL {}{Williams}, A.%
\end{APACrefauthors}%
\unskip\
\newblock
\APACrefYearMonthDay{2021}{{\APACmonth{05}}}{}.
\newblock
{\BBOQ}\APACrefatitle {{A Southern-Hemisphere all-sky radio transient monitor for SKA-Low prototype stations}} {{A Southern-Hemisphere all-sky radio transient monitor for SKA-Low prototype stations}}.{\BBCQ}
\newblock
\APACjournalVolNumPages{Publications of the Astronomical Society of Australia}{38}{}{e023}.
\newblock
\begin{APACrefDOI} \doi{10.1017/pasa.2021.16} \end{APACrefDOI}
\PrintBackRefs{\CurrentBib}

\bibitem [\protect \citeauthoryear {%
{The Pandas Development Team}%
}{%
{The Pandas Development Team}%
}{%
{\protect \APACyear {2023}}%
}]{%
pandas2}
\APACinsertmetastar {%
pandas2}%
\begin{APACrefauthors}%
{The Pandas Development Team}.%
\end{APACrefauthors}%
\unskip\
\newblock
\APACrefYearMonthDay{2023}{{\APACmonth{09}}}{}.
\newblock
\APACrefbtitle {{pandas-dev/pandas: Pandas}.} {{pandas-dev/pandas: Pandas}.}
\newblock
\APAChowpublished {Zenodo}.
\newblock
\APACaddressPublisher{}{Zenodo}.
\newblock
\begin{APACrefDOI} \doi{10.5281/zenodo.3509134} \end{APACrefDOI}
\PrintBackRefs{\CurrentBib}

\bibitem [\protect \citeauthoryear {%
Thekkeppattu%
, Wayth%
\BCBL {}\ \BBA {} Sokolowski%
}{%
Thekkeppattu%
\ \protect \BOthers {.}}{%
{\protect \APACyear {2023}}%
}]{%
zenodo_upload}
\APACinsertmetastar {%
zenodo_upload}%
\begin{APACrefauthors}%
Thekkeppattu, J\BPBI N.%
, Wayth, R\BPBI B.%
\BCBL {}\ \BBA {} Sokolowski, M.%
\end{APACrefauthors}%
\unskip\
\newblock
\APACrefYearMonthDay{2023}{}{}.
\newblock
\APACrefbtitle {Calibration of an SKA-Low prototype station using holographic techniques.} {Calibration of an ska-low prototype station using holographic techniques.}\ [dataset].
\newblock
\begin{APACrefURL} \url{https://doi.org/10.5281/zenodo.8237885} \end{APACrefURL}
\newblock
\begin{APACrefDOI} \doi{10.5281/zenodo.8237885} \end{APACrefDOI}
\PrintBackRefs{\CurrentBib}

\bibitem [\protect \citeauthoryear {%
{van Es}%
\ \protect \BOthers {.}}{%
{van Es}%
\ \protect \BOthers {.}}{%
{\protect \APACyear {2020}}%
}]{%
2020SPIE11445E..89V}
\APACinsertmetastar {%
2020SPIE11445E..89V}%
\begin{APACrefauthors}%
{van Es}, A\BPBI J\BPBI J.%
, {Labate}, M\BPBI G.%
, {Waterson}, M\BPBI F.%
, {Monari}, J.%
, {Bolli}, P.%
, {Davidson}, D.%
\BDBL {}{Paonessa}, F.%
\end{APACrefauthors}%
\unskip\
\newblock
\APACrefYearMonthDay{2020}{{\APACmonth{12}}}{}.
\newblock
{\BBOQ}\APACrefatitle {{A prototype model for evaluating SKA-LOW station calibration}} {{A prototype model for evaluating SKA-LOW station calibration}}.{\BBCQ}
\newblock
\BIn{} H\BPBI K.~{Marshall}, J.~{Spyromilio}\BCBL {}\ \BBA {} T.~{Usuda}\ (\BEDS), \APACrefbtitle {Ground-based and Airborne Telescopes VIII} {Ground-based and airborne telescopes viii}\ (\BVOL\ 11445, \BPG~1144589).
\newblock
\begin{APACrefDOI} \doi{10.1117/12.2562391} \end{APACrefDOI}
\PrintBackRefs{\CurrentBib}

\bibitem [\protect \citeauthoryear {%
{van Haarlem}%
\ \protect \BOthers {.}}{%
{van Haarlem}%
\ \protect \BOthers {.}}{%
{\protect \APACyear {2013}}%
}]{%
2013A&A...556A...2V}
\APACinsertmetastar {%
2013A&A...556A...2V}%
\begin{APACrefauthors}%
{van Haarlem}, M\BPBI P.%
, {Wise}, M\BPBI W.%
, {Gunst}, A\BPBI W.%
, {Heald}, G.%
, {McKean}, J\BPBI P.%
, {Hessels}, J\BPBI W\BPBI T.%
\BDBL {}{van Zwieten}, J.%
\end{APACrefauthors}%
\unskip\
\newblock
\APACrefYearMonthDay{2013}{{\APACmonth{08}}}{}.
\newblock
{\BBOQ}\APACrefatitle {{LOFAR: The LOw-Frequency ARray}} {{LOFAR: The LOw-Frequency ARray}}.{\BBCQ}
\newblock
\APACjournalVolNumPages{Astronomy \& Astrophysics}{556}{}{A2}.
\newblock
\begin{APACrefDOI} \doi{10.1051/0004-6361/201220873} \end{APACrefDOI}
\PrintBackRefs{\CurrentBib}

\bibitem [\protect \citeauthoryear {%
{van Straten}%
, {Jameson}%
\BCBL {}\ \BBA {} {Os\l{}owski}%
}{%
{van Straten}%
\ \protect \BOthers {.}}{%
{\protect \APACyear {2021}}%
}]{%
2021ascl.soft10003V}
\APACinsertmetastar {%
2021ascl.soft10003V}%
\begin{APACrefauthors}%
{van Straten}, W.%
, {Jameson}, A.%
\BCBL {}\ \BBA {} {Os\l{}owski}, S.%
\end{APACrefauthors}%
\unskip\
\newblock
\APACrefYearMonthDay{2021}{{\APACmonth{10}}}{}.
\newblock
\APACrefbtitle {{PSRDADA: Distributed Acquisition and Data Analysis for Radio Astronomy}.} {{PSRDADA: Distributed Acquisition and Data Analysis for Radio Astronomy}.}
\newblock
\APAChowpublished {Astrophysics Source Code Library, record ascl:2110.003}.
\PrintBackRefs{\CurrentBib}

\bibitem [\protect \citeauthoryear {%
Waskom%
}{%
Waskom%
}{%
{\protect \APACyear {2021}}%
}]{%
Waskom2021}
\APACinsertmetastar {%
Waskom2021}%
\begin{APACrefauthors}%
Waskom, M\BPBI L.%
\end{APACrefauthors}%
\unskip\
\newblock
\APACrefYearMonthDay{2021}{}{}.
\newblock
{\BBOQ}\APACrefatitle {seaborn: statistical data visualization} {seaborn: statistical data visualization}.{\BBCQ}
\newblock
\APACjournalVolNumPages{Journal of Open Source Software}{6}{60}{3021}.
\newblock
\begin{APACrefURL} \url{https://doi.org/10.21105/joss.03021} \end{APACrefURL}
\newblock
\begin{APACrefDOI} \doi{10.21105/joss.03021} \end{APACrefDOI}
\PrintBackRefs{\CurrentBib}

\bibitem [\protect \citeauthoryear {%
{Wayth}%
\ \protect \BOthers {.}}{%
{Wayth}%
\ \protect \BOthers {.}}{%
{\protect \APACyear {2022}}%
}]{%
2022JATIS...8a1010W}
\APACinsertmetastar {%
2022JATIS...8a1010W}%
\begin{APACrefauthors}%
{Wayth}, R.%
, {Sokolowski}, M.%
, {Broderick}, J.%
, {Tingay}, S\BPBI J.%
, {Bhushan}, R.%
, {Booler}, T.%
\BDBL {}{Waterson}, M.%
\end{APACrefauthors}%
\unskip\
\newblock
\APACrefYearMonthDay{2022}{{\APACmonth{01}}}{}.
\newblock
{\BBOQ}\APACrefatitle {{Engineering Development Array 2: design, performance, and lessons from an SKA-Low prototype station}} {{Engineering Development Array 2: design, performance, and lessons from an SKA-Low prototype station}}.{\BBCQ}
\newblock
\APACjournalVolNumPages{Journal of Astronomical Telescopes, Instruments, and Systems}{8}{}{011010}.
\newblock
\begin{APACrefDOI} \doi{10.1117/1.JATIS.8.1.011010} \end{APACrefDOI}
\PrintBackRefs{\CurrentBib}

\bibitem [\protect \citeauthoryear {%
{W}es {M}c{K}inney%
}{%
{W}es {M}c{K}inney%
}{%
{\protect \APACyear {2010}}%
}]{%
pandas1}
\APACinsertmetastar {%
pandas1}%
\begin{APACrefauthors}%
{W}es {M}c{K}inney.%
\end{APACrefauthors}%
\unskip\
\newblock
\APACrefYearMonthDay{2010}{}{}.
\newblock
{\BBOQ}\APACrefatitle {{D}ata {S}tructures for {S}tatistical {C}omputing in {P}ython} {{D}ata {S}tructures for {S}tatistical {C}omputing in {P}ython}.{\BBCQ}
\newblock
\BIn{} {S}t\'efan van~der {W}alt\ \BBA {} {J}arrod {M}illman\ (\BEDS), \APACrefbtitle {{P}roceedings of the 9th {P}ython in {S}cience {C}onference} {{P}roceedings of the 9th {P}ython in {S}cience {C}onference}\ (\BPG~56 - 61).
\newblock
\begin{APACrefDOI} \doi{10.25080/Majora-92bf1922-00a} \end{APACrefDOI}
\PrintBackRefs{\CurrentBib}

\bibitem [\protect \citeauthoryear {%
Wijnholds%
}{%
Wijnholds%
}{%
{\protect \APACyear {2017}}%
}]{%
8065418}
\APACinsertmetastar {%
8065418}%
\begin{APACrefauthors}%
Wijnholds, S\BPBI J.%
\end{APACrefauthors}%
\unskip\
\newblock
\APACrefYearMonthDay{2017}{}{}.
\newblock
{\BBOQ}\APACrefatitle {Calibration of Mid-Frequency aperture array stations using self-holography} {Calibration of mid-frequency aperture array stations using self-holography}.{\BBCQ}
\newblock
\BIn{} \APACrefbtitle {2017 International Conference on Electromagnetics in Advanced Applications (ICEAA)} {2017 international conference on electromagnetics in advanced applications (iceaa)}\ (\BPG~967-970).
\newblock
\begin{APACrefDOI} \doi{10.1109/ICEAA.2017.8065418} \end{APACrefDOI}
\PrintBackRefs{\CurrentBib}

\bibitem [\protect \citeauthoryear {%
Wijnholds%
}{%
Wijnholds%
}{%
{\protect \APACyear {2021}}%
}]{%
9560415}
\APACinsertmetastar {%
9560415}%
\begin{APACrefauthors}%
Wijnholds, S\BPBI J.%
\end{APACrefauthors}%
\unskip\
\newblock
\APACrefYearMonthDay{2021}{}{}.
\newblock
{\BBOQ}\APACrefatitle {Generalised Self-Holography} {Generalised self-holography}.{\BBCQ}
\newblock
\BIn{} \APACrefbtitle {2021 XXXIVth General Assembly and Scientific Symposium of the International Union of Radio Science (URSI GASS)} {2021 xxxivth general assembly and scientific symposium of the international union of radio science (ursi gass)}\ (\BPG~1-4).
\newblock
\begin{APACrefDOI} \doi{10.23919/URSIGASS51995.2021.9560415} \end{APACrefDOI}
\PrintBackRefs{\CurrentBib}

\bibitem [\protect \citeauthoryear {%
Wilke%
, Wijnholds%
\BCBL {}\ \BBA {} Gilmore%
}{%
Wilke%
\ \protect \BOthers {.}}{%
{\protect \APACyear {2021}}%
}]{%
9369030}
\APACinsertmetastar {%
9369030}%
\begin{APACrefauthors}%
Wilke, C\BPBI R.%
, Wijnholds, S\BPBI J.%
\BCBL {}\ \BBA {} Gilmore, J.%
\end{APACrefauthors}%
\unskip\
\newblock
\APACrefYearMonthDay{2021}{}{}.
\newblock
{\BBOQ}\APACrefatitle {Calibratability of Aperture Arrays Using Self-Holography} {Calibratability of aperture arrays using self-holography}.{\BBCQ}
\newblock
\APACjournalVolNumPages{IEEE Transactions on Antennas and Propagation}{69}{8}{4527-4537}.
\newblock
\begin{APACrefDOI} \doi{10.1109/TAP.2021.3060070} \end{APACrefDOI}
\PrintBackRefs{\CurrentBib}

\bibitem [\protect \citeauthoryear {%
{Wilke}%
, {Wijnholds}%
\BCBL {}\ \BBA {} {Gilmore}%
}{%
{Wilke}%
\ \protect \BOthers {.}}{%
{\protect \APACyear {2022}}%
}]{%
2022JATIS...8a1008W}
\APACinsertmetastar {%
2022JATIS...8a1008W}%
\begin{APACrefauthors}%
{Wilke}, C\BPBI R.%
, {Wijnholds}, S\BPBI J.%
\BCBL {}\ \BBA {} {Gilmore}, J.%
\end{APACrefauthors}%
\unskip\
\newblock
\APACrefYearMonthDay{2022}{{\APACmonth{01}}}{}.
\newblock
{\BBOQ}\APACrefatitle {{Calibratability of mid-frequency aperture arrays with self-holography}} {{Calibratability of mid-frequency aperture arrays with self-holography}}.{\BBCQ}
\newblock
\APACjournalVolNumPages{Journal of Astronomical Telescopes, Instruments, and Systems}{8}{}{011008}.
\newblock
\begin{APACrefDOI} \doi{10.1117/1.JATIS.8.1.011008} \end{APACrefDOI}
\PrintBackRefs{\CurrentBib}

\end{thebibliography}

%
%
%
%
%

\end{document}